\documentclass[a4paper,11pt]{article}
\pdfoutput=1 

\usepackage{jheppub}

\usepackage{graphicx, epstopdf}

\usepackage[utf8]{inputenc}

\usepackage{amsmath, amsthm, amsfonts, amssymb}
\usepackage{mathtools, mathabx, bm, bbm, scalerel, xfrac}
\usepackage{empheq, physics, slashed, cancel}

\usepackage{hyperref}
\usepackage{color}
\usepackage{xspace} 

\allowdisplaybreaks

\title{Revisiting the double-soft asymptotics of one-loop amplitudes in massless QCD}

\author{Micha\l{} Czakon,}
\author{Felix Eschment and}
\author{Tom Schellenberger}
\affiliation{Institut f\"ur Theoretische Teilchenphysik und Kosmologie, RWTH Aachen University,\\ D-52056 Aachen, Germany}
\emailAdd{mczakon@physik.rwth-aachen.de}
\emailAdd{felix.eschment@rwth-aachen.de}
\emailAdd{tom.schellenberger@rwth-aachen.de}

\abstract{We evaluate the one-loop soft current for the emission of two soft gluons or a soft quark-anti-quark pair in massless Quantum Chromodynamics. The results are exact in dimensional regularisation up to a single Feynman integral. Two terms of the Taylor series of the latter integral as a function of $\epsilon \equiv (4-d)/2$ with $d$ the dimension of spacetime are available from a recent calculation of one-loop triple-collinear splitting functions. Our formulae are necessary for the construction of a subtraction scheme for the evaluation of next-to-next-to-next-to-leading order cross sections in massless QCD.}
\keywords{QCD, Scattering Amplitudes, Higher-Order Perturbative Calculations}
\preprint{P3H-22-111, TTK-22-37}

\begin{document}
\maketitle
\flushbottom

\section{Introduction}

Current research aiming at the development of efficient methods for cross section evaluation at next-to-next-to-next-to-leading order (N3LO) in QCD requires complete knowledge of the singular asymptotics of scattering amplitudes in relevant soft and collinear limits. The collinear splitting operators sufficient for N3LO applications are available from Refs.~\cite{DelDuca:1999iql, Birthwright:2005ak, Birthwright:2005vi, DelDuca:2019ggv, DelDuca:2020vst, Catani:2003vu, Sborlini:2014mpa, Sborlini:2014kla, Badger:2015cxa, Czakon:2022fqi, Bern:2004cz, Badger:2004uk, Duhr:2014nda, Catani:2011st}. The indispensable soft asymptotics of massless QCD, on the other hand, has been analysed for the emission of either three soft gluons or a soft gluon together with a soft quark-anti-quark pair at the tree level in Refs.~\cite{Catani:2019nqv, DelDuca:2022noh, Catani:2022hkb}, and for the emission of a single soft gluon at the two-loop level in Refs.~\cite{Badger:2004uk, Li:2013lsa, Duhr:2013msa, Dixon:2019lnw}.

The emission of either two soft gluons or a soft quark-anti-quark pair at the one-loop level, the subject of the present publication, has been studied previously in Refs.~\cite{Zhu:2020ftr,Catani:2021kcy}. However, both references only provide formulae for Laurent-series terms that do not vanish in the four-dimensional limit of dimensional regularisation. While this allows to regulate N3LO phase-space integrals involving one-loop amplitudes with the help of a subtraction of the singular asymptotics, it is nevertheless insufficient to obtain the necessary integral of the latter subtraction term, and thus the complete contribution of real radiation at the one-loop level. Our goal is to remedy this shortcoming. In particular, we provide the soft current in massless QCD for double soft-gluon and soft quark-anti-quark pair emission with exact dependence on the spacetime dimension $d \equiv 4-2\epsilon$ in dimensional regularisation. The results are expressed in terms of several one-loop master integrals, all of which but one have been evaluated in Ref.~\cite{Zhu:2020ftr} as exact functions of $d$. The one integral that does not admit a closed-form analytic expression has been recently evaluated to $\order{\epsilon}$ in Ref.~\cite{Czakon:2022fqi}. This result is sufficient for N3LO applications in massless QCD. On the other hand, processes with massive partons require a further generalisation that has not been attempted in the literature yet. Finally, we remind that calculations at N3LO also hinge on soft/collinear asymptotics of amplitudes at lower orders. Fortunately, these are available from Refs.~\cite{Campbell:1997hg, Catani:1998nv, Bern:1998sc, Kosower:1999rx, Bern:1999ry, Catani:1999ss, Catani:2000pi, Czakon:2011ve, Bierenbaum:2011gg, Czakon:2018iev, Catani:2011st, Sborlini:2013jba}.

The paper is organised as follows. In the next section, we define the soft current for multiple soft-gluon emission. Subsequently, in Section~\ref{sec:Diagrams}, we discuss the occurring one-loop diagrams in the case of two soft gluons from the perspective of colour and spin dependence. This leads to a decomposition of the result into several scalar functions. The latter are provided in Section~\ref{sec:ExactResults} in terms of one-loop master integrals retaining the exact dependence on $d$. Expanded versions of the same functions in $\epsilon$ up to $\order{\epsilon^0}$ are the subject of Section~\ref{sec:ExpandedResults}. Section~\ref{sec:SoftQuarks} contains analogous results for a soft quark-anti-quark pair. We conclude with an outlook on future work on the soft asymptotics. The main text is complemented with four appendices. Appendix~\ref{app:Integrals} provides the results for the one-loop master integrals. In Appendix~\ref{app:Tests}, we present several numerical tests that we have performed in order to verify the correctness of our results. Appendix~\ref{app:Erratum} lists typos that we have found in Ref.~\cite{Zhu:2020ftr}. The latter publication is valuable, since it contains the double-soft asymptotics in the spinor helicity formalism as well as a discussion of iterated soft limits. Hence, we believe that it is useful to point out minor mistakes in the formulae to save the interested reader from the burden of recalculating everything from scratch. This is an added bonus of our study. Finally, Appendix~\ref{app:Files} describes the results that we provide in electronic form.

\section{The soft current for multiple soft-gluon emission}

Let $\ket{M(q_1,\dots,q_n,\{ p_i \}^m_{i=1})}$ be the amplitude for a scattering process with $n$ outgoing (soft) gluons with momenta $q_1,\dots, q_n$, $m$ arbitrary (hard) massless partons with momenta $p_i$, $p_i^2 = 0$, and an arbitrary number of colour-neutral particles whose momenta have been omitted as irrelevant to the problem. The amplitude is defined as a vector in colour and spin space for each configuration of the colour and spin quantum numbers (for more details see for example Ref.~\cite{Catani:1996vz}). Its leading asymptotics in the soft limit of vanishing momenta of the $n$ gluons, $q_i \to 0$, $i = 1,\dots,n$, is given by:
\begin{multline} \label{eq:asymptotics}
    \ip{a_1\lambda_1,\dots,a_n\lambda_n}{M(q_1,\dots,q_n,\{ p_i \}^m_{i=1})} \quad \sim \\[.2cm] \epsilon_{\lambda_1}^{\alpha_1\,*}(q_1) \dots \epsilon_{\lambda_n}^{\alpha_n\,*}(q_n) \, \bm{\mathrm{J}}^{a_1\dots a_n}_{\alpha_1\dots\alpha_n}\big( q_1,\dots,q_n,\{ p_i\}^m_{i=1}\big) \ket{M(\{ p_i\}^m_{i=1})} \; ,
\end{multline}
where $a_i$ and $\lambda_i$ are adjoint-representation colour indices and helicities of the soft gluons, while $\epsilon_{\lambda_i}^{\alpha_i\,*}(q_i)$ are the respective polarisation vectors. The reduced amplitude $\ket{M(\{ p_i\}^m_{i=1})}$ corresponds to the scattering of the same partons and colour-neutral particles as $\ket{M(q_1,\dots,q_n,\{ p_i \}^m_{i=1})}$ but without the soft gluons. The asymptotics is valid in the general soft limit without a strong hierarchy of softness, i.e.\  for $q_i^0/q_j^0 \to \kappa_{ij} > 0$, as well as in the case of a hierarchy of softness, called iterated soft limit, where some $\kappa_{ij} \to 0$. The soft current $\bm{\mathrm{J}}^{a_1\dots a_n}_{\alpha_1\dots\alpha_n}\big( q_1,\dots,q_n,\{ p_i\}^m_{i=1}\big)$ is an operator in colour space that is trivial in spin space, and does not depend on the momenta of colour-neutral particles. It is symmetric w.r.t.\  the exchange of the parameters of any two soft gluons. Furthermore, as a consequence of the difference in the mass dimension of the amplitudes on the left- and right-hand side of Eq.~\eqref{eq:asymptotics} with dimensionless polarisation vectors, it has the mass dimension:
\begin{equation} \label{eq:Jdim}
    \dim \bm{\mathrm{J}}^{a_1\dots a_n}_{\alpha_1\dots\alpha_n}\big( q_1,\dots,q_n,\{ p_i\}^m_{i=1}\big) = -n ( 1-\epsilon) \; ,
\end{equation}
where we have assumed dimensional regularisation in $d = 4-2\epsilon$ spacetime dimensions.

The soft current can be obtained from a matrix element of a time-ordered product of Wilson-line operators, $W_i(x)$ (for a pedagogical discussion see for example Ref.~\cite{Feige:2014wja}), corresponding to the eikonal approximation:
\begin{equation} \label{eq:JbyWilson}
    \epsilon_{\lambda_1}^{\alpha_1\,*}(q_1) \dots \epsilon_{\lambda_n}^{\alpha_n\,*}(q_n) \, \bm{\mathrm{J}}^{a_1\dots a_n}_{\alpha_1\dots\alpha_n}\big( q_1,\dots,q_n,\{ p_i\}^m_{i=1}\big) = \mel{a_1q_1\lambda_1,\dots, a_nq_n\lambda_n}{T\bigg[\prod_{i=1}^m W_i(0)\bigg]}{\Omega} \; .
\end{equation}
Specifically, if $i$ is an outgoing parton, then:
\begin{equation} \label{eq:Wilson}
    W_i(x) \equiv \lim_{\; \; \eta \to 0^+} P\exp(ig^B_s \int_0^\infty \dd{t} e^{-\eta t} \, \beta^\alpha_i A_\alpha^a(x + \beta_i t) \, \bm{\mathrm{T}}_i^a ) \; , \qquad \beta^\alpha_i \equiv p^\alpha_i/p_i^0 \; ,
\end{equation}
where $g^B_s$ is the bare strong coupling constant. We do not renormalise the gluon field $A^a_\alpha(\cdot)$, since without renormalisation, the LSZ normalisation constants in massless QCD vanish. If we did renormalise the gluon field, then the Wilson-line operator would be defined with the bare field, and we would have to include non-trivial LSZ normalisation constants in Eq.~\eqref{eq:JbyWilson}. The colour-charge operator $\bm{\mathrm{T}}^a_i \equiv \sum_{bc} \ket{b}\mel{b}{\bm{\mathrm{T}}_i^a}{c}\bra{c}$ acts non-trivially only on the colour index $c$ of parton $i$ in the reduced amplitude $\ket{M(\{ p_i \}^m_{i=1})}$. In particular, for a gluon there is $\mel{b}{\bm{\mathrm{T}}_i^a}{c} = if^{bac}$, for an outgoing quark $\mel{b}{\bm{\mathrm{T}}_i^a}{c} = T^a_{bc}$, whereas for an outgoing anti-quark $\mel{b}{\bm{\mathrm{T}}_i^a}{c} = -T^a_{cb}$. As usual, the structure constants $f^{abc}$ are defined by $\comm{\bm{\mathrm{T}}^a_i}{\bm{\mathrm{T}}^b_j} = i f^{abc} \bm{\mathrm{T}}^a_i \delta_{ij}$, while the fundamental-representation generators, $T^a_{bc}$, are normalised with $\Tr(T^a T^b) = T_F \delta^{ab}$. Later, we will also encounter the two Casimirs, $C_A = 2T_F N_c$ for the adjoint representation, and $C_F = T_F (N_c^2-1)/N_c$ for the fundamental representation, with $N_c$ the number of colours for the SU$(N_c)$ gauge group, and in particular $N_c = 3$ for QCD. Finally, $P\exp(\cdot)$ is a path ordered exponential that orders the colour-charge operators along the integration path from right to left.

Eq.~\eqref{eq:Wilson} corresponds to the classical motion with ``time'' $t$ and ``velocity'' $\beta_i$ of the massless hard parton $i$ from the spacetime point $x$ to $x + \beta_i t$, $t \to \infty$. Any rescaling of $\beta_i$ with a positive constant can be absorbed in the integration measure. The regularisation with an exponential, $e^{-\eta t}$, is necessary to assure convergence. The term in the expansion of $W_i(0)$ corresponding to the emission of $n$ outgoing gluons, either real or virtual, with momenta $k_j$, Lorentz indices $\alpha_j$, and colour indices $a_j$, is:
\begin{equation} \label{eq:WilsonExp}
\frac{-g_s^B p_i^{\alpha_1} \bm{\mathrm{T}}_i^{a_1}}{p_i \cdot k_1 + i0^+}  \frac{-g_s^B p_i^{\alpha_2} \bm{\mathrm{T}}_i^{a_2}}{p_i \cdot ( k_1 + k_2 ) + i0^+} \cdots \frac{-g_s^B p_i^{\alpha_n} \bm{\mathrm{T}}_i^{a_n}}{p_i \cdot ( k_1 + k_2 + \dots + k_n ) + i0^+} + \text{gluon permutations} \; .
\end{equation}
Feynman's causal prescription, $+i0^+$, for eikonal propagators, $1/(p_i \cdot \dots)$, is only relevant for virtual emissions.

The Wilson-line operator of an incoming parton $i$ is defined as follows: 
\begin{equation} \label{eq:WilsonIncoming}
    W_i(x) \equiv \lim_{\; \; \eta \to 0^+} P\exp(ig_s^B \int_{-\infty}^0 \dd{t} e^{\eta t} \, \beta^\alpha_i \cdot A_\alpha^a(x + \beta_i t) \, \bm{\mathrm{T}}_i^a ) \; ,
\end{equation}
where the colour-charge operator is the same as that of the corresponding outgoing parton. In momentum space, this yields Eq.~\eqref{eq:WilsonExp} with the replacements $+i0^+ \to -i0^+$ and $\mel{b}{\bm{\mathrm{T}}_i^a}{c} \to - \mel{c}{\bm{\mathrm{T}}_i^a}{b}$. We consolidate the treatment of outgoing and incoming partons by using different colour-charge operators for incoming partons according to this replacement. This is also the standard approach adapted e.g.\ in Ref.~\cite{Catani:1996vz}.

Due to the dependence of Eq.~\eqref{eq:Wilson} on $\beta_i$ rather than $p_i$, the soft current is rescaling invariant w.r.t.\  the momenta of the hard partons:
\begin{equation} \label{eq:rescaling}
    \bm{\mathrm{J}}^{a_1\dots a_n}_{\alpha_1\dots\alpha_n}\big( q_1,\dots,q_n,\{ \alpha_i p_i\}^m_{i=1}\big) = \bm{\mathrm{J}}^{a_1\dots a_n}_{\alpha_1\dots\alpha_n}\big( q_1,\dots,q_n,\{ p_i\}^m_{i=1}\big) \; , \qquad \alpha_i > 0 \; .
\end{equation}
The same conclusion follows from inspection of Eq.~\eqref{eq:WilsonExp}. The QCD Ward identity implies that for any $k$:
\begin{equation} \label{eq:Ward}
    \eval{\epsilon_{\lambda_1}^{\alpha_1\,*}(q_1) \dots \epsilon_{\lambda_n}^{\alpha_n\,*}(q_n) \, \bm{\mathrm{J}}^{a_1\dots a_n}_{\alpha_1\dots\alpha_n}\big( q_1,\dots,q_n,\{ p_i\}^m_{i=1}\big) \; }_{\epsilon_{\lambda_k}^{\alpha_k\,*}(q_k) = q_k^{\alpha_k}} \quad \propto \quad \sum_{i=1}^m \bm{\mathrm{T}}_i^{a_k} \; ,
\end{equation}
which vanishes by colour conservation when acting on a scattering amplitude.

The soft current for $n$ soft gluons admits a perturbative loop expansion:
\begin{equation} \label{eq:Jexp}
    \bm{\mathrm{J}} = \big( g_s^B \big)^n \bigg( \bm{\mathrm{J}}^{(0)} + \frac{\mu^{-2\epsilon} \alpha_s^B}{(4\pi)^{1-\epsilon}} \, \bm{\mathrm{J}}^{(1)} + \dots \bigg) \; , \qquad \alpha_s^B \equiv \frac{\big( g_s^B \big)^2}{4\pi} \; ,
\end{equation}
where we have only kept explicit the tree-level, $\bm{\mathrm{J}}^{(0)}$, and one-loop, $\bm{\mathrm{J}}^{(1)}$, contributions that are of interest for the present publication. Furthermore, we have suppressed colour and Lorentz indices as well as the momenta of the involved partons. As defined above, the one-loop soft current is not renormalised. Furthermore, in view of Eqs.~\eqref{eq:Jdim} and \eqref{eq:rescaling}, both the tree-level and one-loop soft currents have an integer mass dimension due entirely to the soft momenta and the unit of mass $\mu$:
\begin{align}
    &\bm{\mathrm{J}}^{(0)}{}^{a_1\dots a_n}_{\alpha_1\dots\alpha_n}\big( \lambda q_1,\dots,\lambda q_n,\{ p_i\}^m_{i=1}\big) =& &\lambda^{-n} \, \bm{\mathrm{J}}^{(0)}{}^{a_1\dots a_n}_{\alpha_1\dots\alpha_n}\big( q_1,\dots,q_n,\{ p_i\}^m_{i=1}\big) \; , &\qquad \lambda > 0 \; , \label{eq:Jdim0} \\[.2cm] \label{eq:Jdim1}
    &\bm{\mathrm{J}}^{(1)}{}^{a_1\dots a_n}_{\alpha_1\dots\alpha_n}\big( \lambda \mu, \lambda q_1,\dots,\lambda q_n,\{ p_i\}^m_{i=1}\big) =& &\lambda^{-n} \, \bm{\mathrm{J}}^{(1)}{}^{a_1\dots a_n}_{\alpha_1\dots\alpha_n}\big( \mu, q_1,\dots,q_n,\{ p_i\}^m_{i=1}\big) \; . &
\end{align}

In the case of a single soft-gluon emission, Eq.~\eqref{eq:WilsonExp} yields the tree-level soft current:
\begin{equation} \label{eq:1gluonJ0}
    \bm{\mathrm{J}}^{(0)}{}^{a}_{\alpha}\big( q,\{ p_i \}_{i=1}^m \big) = - \sum_{i} \bm{\mathrm{T}}_i^a \frac{p_{i\alpha}}{p_i \cdot q} \; .
\end{equation}
The one-loop soft current, on the other hand, is given by \cite{Catani:2000pi}:
\begin{equation} \label{eq:1gluonJ1}
    \bm{\mathrm{J}}^{(1)}{}^{a}_{\alpha}\big( q,\{ p_i \}_{i=1}^m \big) = r_\Gamma \frac{\Gamma(1-\epsilon)\Gamma(1+\epsilon)}{\epsilon^2} \, if^{abc} \sum_{i \neq j} \bm{\mathrm{T}}_i^b \bm{\mathrm{T}}_j^c \, \bigg( \frac{p_{i\alpha}}{p_i \cdot q} - \frac{p_{j\alpha}}{p_j \cdot q} \bigg) \bigg( \frac{\mu^2 s_{ij}}{s_{qi} s_{qj}} \bigg)^{\epsilon} e^{i\pi \epsilon \sigma_{ij}} \; ,
\end{equation}
with $s_{ij} \equiv 2p_i \cdot p_j$, $s_{qi} \equiv 2 q \cdot p_i$, $s_{qj} \equiv 2 q \cdot p_j$, and $\sigma_{ij} = -1$ for both $i$ and $j$ incoming, and $\sigma_{ij} = +1$ otherwise. We have also defined the following constant for future reference:
\begin{equation} \label{eq:rGamma}
    r_\Gamma \equiv \frac{\Gamma^2(1-\epsilon)\Gamma(1+\epsilon)}{\Gamma(1-2\epsilon)} \; .
\end{equation}
Notice that:
\begin{equation} \label{eq:1gluonJ0Ward}
    q \cdot \bm{\mathrm{J}}^{(0)\,a}\big( q,\{ p_i \}_{i=1}^m \big) = - \sum_{i} \bm{\mathrm{T}}_i^a \; ,
\end{equation}
while:
\begin{equation}\label{eq:1gluonJ1Ward}
    q \cdot \bm{\mathrm{J}}^{(1)\,a}\big( q,\{ p_i \}_{i=1}^m \big) = 0 \; .
\end{equation}
The last relation follows from the structure of the contents of the first bracket in the summand in Eq.~\eqref{eq:1gluonJ1}, which itself is a necessary consequence of the fact that the said summand must be proportional to $\big( \mu^2 s_{ij} / s_{qi} s_{qj} \big)^\epsilon$ by Eqs.~\eqref{eq:rescaling}, \eqref{eq:Jexp} and \eqref{eq:Jdim1}. Let us also define:
\begin{equation} \label{eq:1gluonJ0alt}
\begin{split}
    \bm{\mathcal{J}}^{(0)}{}^{a}_{\alpha}\big( q,\{ p_i \}_{i=1}^m \big) &\equiv - \frac{1}{C_A} if^{abc} \sum_{i \neq j} \bm{\mathrm{T}}_i^b \bm{\mathrm{T}}_j^c \, \bigg( \frac{p_{i\alpha}}{p_i \cdot q} - \frac{p_{j\alpha}}{p_j \cdot q} \bigg) \\[.2cm]
    &= \bm{\mathrm{J}}^{(0)}{}^{a}_{\alpha}\big( q,\{ p_i \}_{i=1}^m \big) - \frac{2}{C_A} if^{abc} \sum_{i} \bm{\mathrm{T}}_i^b \, \frac{p_{i\alpha}}{p_i \cdot q} \sum_{j} \bm{\mathrm{T}}_j^c\; .
\end{split}
\end{equation}
$\bm{\mathcal{J}}^{(0)}{}^{a}_{\alpha}$ trivially satisfies the Ward identity just as $\bm{\mathrm{J}}^{(1)}{}^{a}_{\alpha}$. Furthermore, the second term on the r.h.s.\ of Eq.~\eqref{eq:1gluonJ0alt} vanishes by colour conservation.

Before discussing the main topic of the present publication in the next section, we also reproduce the tree-level soft current for the emission of two soft gluons \cite{Berends:1988zn,Catani:1999ss}:
\begin{multline} \label{eq:2gluonJ0}
    \bm{\mathrm{J}}^{(0)}{}^{a_1a_2}_{\alpha_1\alpha_2}\big( q_1,q_2 \big) = \frac{1}{2} \, \acomm{\bm{\mathrm{J}}^{(0)}{}^{a_1}_{\alpha_1}\big( q_1 \big)}{\bm{\mathrm{J}}^{(0)}{}^{a_2}_{\alpha_2}\big( q_2 \big)} \\[.2cm] + i f^{a_1a_2c} \sum_i \bm{\mathrm{T}}_i^c \Bigg[ \frac{p_{i\alpha_1} q_{1\alpha_2} - p_{i\alpha_2} q_{2\alpha_1}}{(q_1 \cdot q_2) \big( p_i \cdot (q_1 + q_2)\big)} - \frac{p_i \cdot (q_1 - q_2)}{2 p_i \cdot ( q_1 + q_2 )} \bigg( \frac{p_{i\alpha_1} p_{i\alpha_2}}{(p_i \cdot q_1) (p_i \cdot q_2)} + \frac{g_{\alpha_1\alpha_2}}{q_1 \cdot q_2} \bigg) \Bigg] \; ,
\end{multline}
where we have suppressed the momenta of the hard partons in the argument list of the soft currents, as we shall do from now on. The anti-commutator of single soft-gluon emission soft currents on the first line of the r.h.s.\ of Eq.~\eqref{eq:2gluonJ0} covers the case of independent emissions by two different hard partons. However, it also contains the abelian part of the emission by the same parton, where the colour-charge operators are effectively commuting, because the product of two colour-charge operators acting on the same line is replaced by their anti-commutator (see Eq.~\eqref{eq:eikonalIdentity} below). Hence, the second line of Eq.~\eqref{eq:2gluonJ0} is purely non-abelian. On the other hand, because the tree-level current for single soft-gluon emission does not vanish identically when contracted with the soft-gluon momentum, Eq.~\eqref{eq:1gluonJ0Ward}, the terms on the first and second lines of Eq.~\eqref{eq:2gluonJ0} do not satisfy the Ward identity \eqref{eq:Ward} separately. Nevertheless, the identity is satisfied by the sum:
\begin{equation} \label{eq:2gluonJ0Ward}
    q_1^{\alpha_1} \bm{\mathrm{J}}^{(0)}{}^{a_1a_2}_{\alpha_1\alpha_2} \big( q_1,q_2 \big) = \bigg( - \delta^{a_1 c} \bm{\mathrm{J}}^{(0)}{}^{a_2}_{\alpha_2}\big( q_2 \big) + \frac{1}{2} \frac{q_1{}_{\alpha_2}}{q_1 \cdot q_2} i f^{a_1a_2 c} \bigg) \sum_{i} \bm{\mathrm{T}}_i^c \; .
\end{equation}

\section{Double soft-gluon emission}

\subsection{Structure in colour and spin space} \label{sec:Diagrams}

\begin{figure}
\begin{minipage}{.45\textwidth}
    \begin{center}
    \includegraphics[scale=.7, page=1]{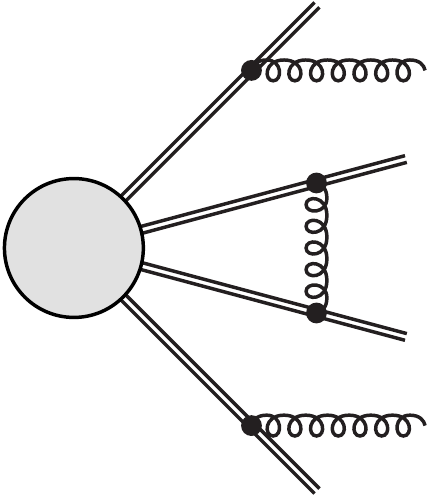}
    \caption{Double soft-gluon emission diagram involving four Wilson lines. The loop integral is scaleless and vanishes. \label{fig:4lines}}
    \end{center}
    \vspace{1cm}
\end{minipage}
\hspace{.5cm}
\begin{minipage}{.45\textwidth}
    \begin{center}
    \includegraphics[scale=.7, page=1]{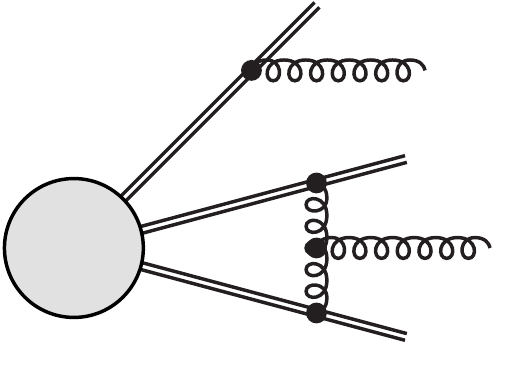}
    \caption{Double soft-gluon emission diagram involving three Wilson lines. The contribution is a product of tree-level and one-loop single soft-gluon emissions. \label{fig:3lines}}
    \end{center}
\end{minipage}
\end{figure}

Let $W^{(n)}_i$ be the order-$n$ truncation of the formal Taylor expansion in $g_s^B$ of the Wilson-line operator $W_i$ defined in Eq.~\eqref{eq:Wilson} for outgoing and in Eq.~\eqref{eq:WilsonIncoming} for incoming partons:
\begin{equation}
W_i \equiv W^{(n)}_i + \order{\big(g^B_s\big)^{n+1}} \; , \qquad W^{(0)}_i = \mathbbm{1} \; .
\end{equation}
According to Eq.~\eqref{eq:JbyWilson}, the one-loop soft current for double soft-gluon emission, $\mathbf{J}^{(1)}{}^{a_1 a_2}_{\alpha_1 \alpha_2}(q_1, q_2)$, defined by Eq.~\eqref{eq:Jexp} with $n = 2$, contains terms stemming from products $W^{(n_1)}_{i_1} \cdot W^{(n_2)}_{i_2} \cdot \dots \cdot W^{(n_W)}_{i_W}$ with $n_i > 0$ and $n_1 + \dots + n_W = n + 2 - n_g$, where $n_g$ is the power of $g_s^B$ due to QCD interactions. The only diagram generated by four non-trivial Wilson lines is depicted in Fig.~\ref{fig:4lines}. The loop integral in this diagram is scaleless, which implies that this contribution vanishes. An example diagram generated by three non-trivial Wilson lines is depicted in Fig.~\ref{fig:3lines}. All diagrams of this type are entirely described by the product of one-loop and tree-level soft currents for single soft-gluon emissions. In fact, the diagram of Fig.~\ref{fig:3lines} is the only one that does not involve a scaleless integral and is thus the only one that contributes. We thus define the non-trivial part of the double soft-gluon emission current, $\Delta \mathbf{J}^{(1)}{}^{a_1 a_2}_{\alpha_1 \alpha_2}(q_1, q_2)$, as follows:
\begin{empheq}[box=\fbox]{equation} \label{eq:DeltaJ1def}
\mathbf{J}^{(1)}{}^{a_1 a_2}_{\alpha_1 \alpha_2}(q_1, q_2) = \left( \mathbf{J}^{(1)}{}^{a_1}_{\alpha_1}(q_1) \mathbf{J}^{(0)}{}^{a_2}_{\alpha_2}(q_2) + \mathbf{J}^{(1)}{}^{a_2}_{\alpha_2}(q_2) \mathbf{J}^{(0)}{}^{a_1}_{\alpha_1}(q_1) \right) + \Delta \mathbf{J}^{(1)}{}^{a_1 a_2}_{\alpha_1 \alpha_2}(q_1, q_2) \; .
\end{empheq}
In principle, one could proceed as in the case of the tree-level soft current, Eq.~\eqref{eq:2gluonJ0}, and use anti-commutators instead of simple products,  $\mathbf{J}^{(1)}\mathbf{J}^{(0)} \; \to \; 1/2 \, \acomm{\mathbf{J}^{(1)}}{\mathbf{J}^{(0)}}$. Then, however, $\Delta \mathbf{J}^{(1)}{}^{a_1 a_2}_{\alpha_1 \alpha_2}(q_1, q_2)$ would not satisfy the Ward identity \eqref{eq:Ward} on its own. Indeed, while the r.h.s.\ of Eq.~\eqref{eq:DeltaJ1def} must have the form $(\dots) \sum_i \bm{\mathrm{T}}_i^{a_1}$ after contraction with $q_1^{\alpha_1}$ for example, a definition with anti-commutators would generate the term $1/2 \, q_1^{\alpha_1} \mathbf{J}^{(0)}{}^{a_1}_{\alpha_1}(q_1) \mathbf{J}^{(1)}{}^{a_2}_{\alpha_2}(q_2) = - 1/2 \sum_{i} \bm{\mathrm{T}}_i^{a_1} \mathbf{J}^{(1)}{}^{a_2}_{\alpha_2}(q_2) \; \neq \; (\dots) \sum_i \bm{\mathrm{T}}_i^{a_1}$. This term would have to be cancelled by an appropriate contribution from $q_1^{\alpha_1} \Delta \mathbf{J}^{(1)}{}^{a_1 a_2}_{\alpha_1 \alpha_2}(q_1,q_2)$. In view of the complexity of the expression for the one-loop soft current, it is desirable to enforce and exploit as many constraints as possible. The Ward identity is thus a welcome property, and we retain Eq.~\eqref{eq:DeltaJ1def}.

\begin{figure}
\begin{tabular}{cccc}
    \includegraphics[scale=0.7, page=1]{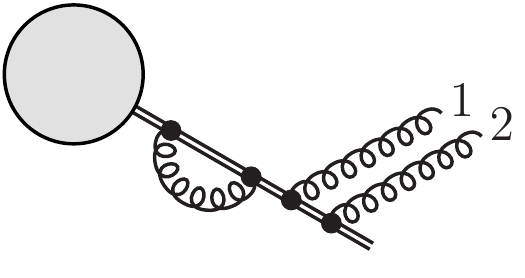}&
    \includegraphics[scale=0.7, page=2]{diagrams/diagrams-crop.pdf}&
    \includegraphics[scale=0.7, page=3]{diagrams/diagrams-crop.pdf}&    
    \includegraphics[scale=0.7, page=4]{diagrams/diagrams-crop.pdf} \\    
    \includegraphics[scale=0.7, page=5]{diagrams/diagrams-crop.pdf}&    
    \includegraphics[scale=0.7, page=6]{diagrams/diagrams-crop.pdf}&    
    \includegraphics[scale=0.7, page=7]{diagrams/diagrams-crop.pdf}&    
    \includegraphics[scale=0.7, page=8]{diagrams/diagrams-crop.pdf} \\    
    \includegraphics[scale=0.7, page=9]{diagrams/diagrams-crop.pdf}&    
    \includegraphics[scale=0.7, page=10]{diagrams/diagrams-crop.pdf}&    
    \includegraphics[scale=0.7, page=11]{diagrams/diagrams-crop.pdf}&    
    \includegraphics[scale=0.7, page=12]{diagrams/diagrams-crop.pdf} \\   
    \includegraphics[scale=0.7, page=13]{diagrams/diagrams-crop.pdf}&    
    \includegraphics[scale=0.7, page=14]{diagrams/diagrams-crop.pdf}&    
    \includegraphics[scale=0.7, page=15]{diagrams/diagrams-crop.pdf}&    
    \includegraphics[scale=0.7, page=16]{diagrams/diagrams-crop.pdf} \\ 
    \includegraphics[scale=0.7, page=32]{diagrams/diagrams-crop.pdf}&
    \includegraphics[scale=0.7, page=33]{diagrams/diagrams-crop.pdf}&    
\end{tabular}
    \caption{Diagrams with two soft-gluon emissions from a single Wilson line (denoted by a double line) that yield scaleless integrals assuming that the emitting hard parton is massless. Numbers 1 and 2 have been inserted when the value of a given diagram depends on the order of emissions. The shaded circle represents the reduced amplitude on which the soft current acts through colour-charge operators. \label{fig:scaleless1line}}
\end{figure}

\begin{figure}
\begin{tabular}{cccc}
    \includegraphics[scale=0.7, page=21]{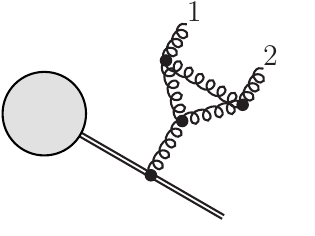}&    
    \includegraphics[scale=0.7, page=22]{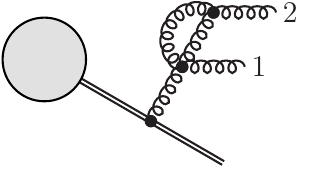}&    
    \includegraphics[scale=0.7, page=23]{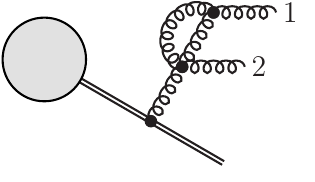}&    
    \includegraphics[scale=0.7, page=24]{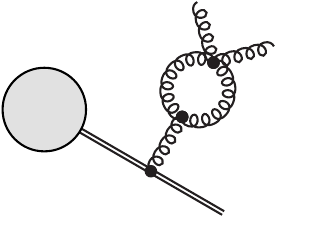} \\   
    \includegraphics[scale=0.7, page=25]{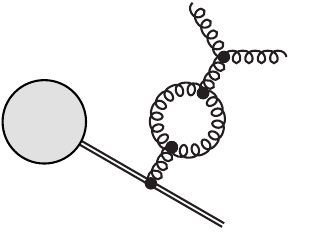}&    
    \includegraphics[scale=0.7, page=26]{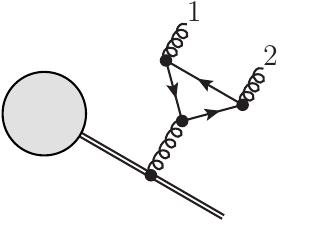}&    
    \includegraphics[scale=0.7, page=27]{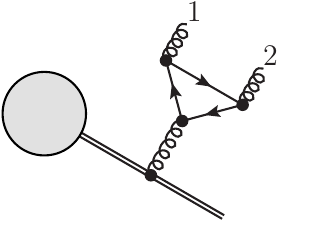}&    
    \includegraphics[scale=0.7, page=28]{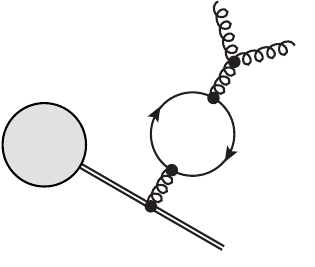}\\
    \includegraphics[scale=0.7, page=29]{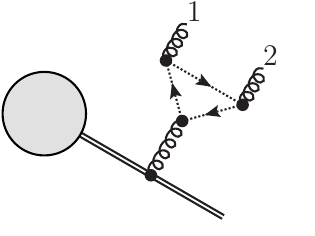}&    
    \includegraphics[scale=0.7, page=30]{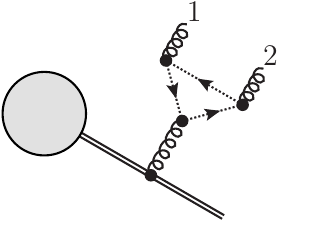}&    
    \includegraphics[scale=0.7, page=31]{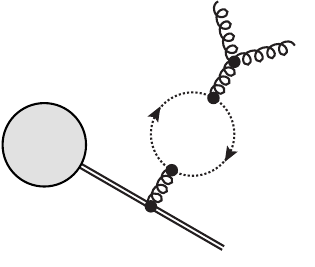}& 
\end{tabular}
    \caption{Diagrams with a single off-shell gluon coupling to the Wilson line. Description of the diagrams as in Fig.~\ref{fig:scaleless1line}. Notice the presence of a quark loop (denoted by a single line with arrow for fermion flow direction) as well as that of a ghost loop (denoted by a dotted line). Diagrams with ghosts are due to the use of the Feynman gauge. \label{fig:1eikonal1line}}
\end{figure}

\begin{figure}
\begin{tabular}{cccc}
    \includegraphics[scale=0.7, page=17]{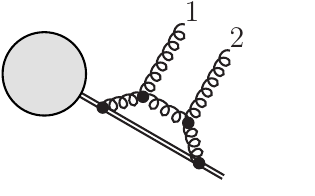}&    
    \includegraphics[scale=0.7, page=18]{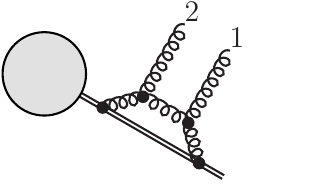}&    
    \includegraphics[scale=0.7, page=19]{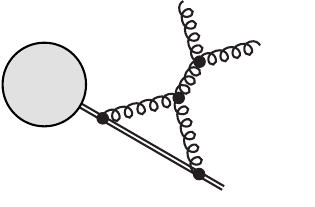}&    
    \includegraphics[scale=0.7, page=20]{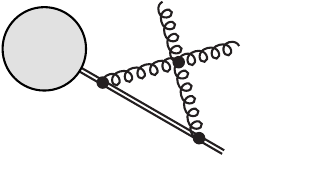}\\ 
\end{tabular}
    \caption{Remaining diagrams with two soft-gluon emissions from a single Wilson line. Description as in Fig.~\ref{fig:scaleless1line}. \label{fig:main1line}}
\end{figure}

After taking care of contributions involving three different Wilson lines, there still remain possible contributions with emissions from a single or from two different Wilson lines. Let us consider the former first. The contributing diagrams are depicted in Figs.~\ref{fig:scaleless1line}, \ref{fig:1eikonal1line} and \ref{fig:main1line}. Since $\Delta \mathbf{J}^{(1)}$ has integer dimension, but is proportional to $\mu^{2\epsilon}$, see Eq.~\eqref{eq:Jexp}, and rescaling invariant w.r.t.\  the hard momentum of the Wilson line, $p_i$, see Eq.~\eqref{eq:rescaling}, the value of any diagram with a single emitting Wilson must be proportional to:
\begin{equation} \label{eq:prefactor1}
    \bigg( \frac{\mu^2}{q_1 \cdot q_2} \bigg)^{\epsilon} \; .
\end{equation}
Hence, if the Feynman integral corresponding to a given diagram does not depend on $q_1 \cdot q_2$, the result must vanish (scaleless diagram/integral). In order to identify such integrals, we note that if a soft-gluon couples directly to the Wilson line, the dependence on its momentum $q_j$ with $j = 1$ or $j = 2$, is only through the scalar product $p_i \cdot q_j$. Indeed, any eikonal propagator depending on $q_j$ has the form:
\begin{equation}
    \frac{1}{p_i \cdot (k + q_j)} = \frac{1}{(p_i \cdot k) + (p_i \cdot q_j)} \; ,
\end{equation}
where $k$ is a linear combination of the loop momentum (possibly with vanishing coefficient) and the second soft-gluon momentum (possibly with vanishing coefficient). In this case there is thus no dependence on $q_1 \cdot q_2$. This argument identifies all diagrams in Fig.~\ref{fig:scaleless1line} as scaleless and hence vanishing.

Fig.~\ref{fig:1eikonal1line} contains all diagrams where the coupling of the soft gluons to the Wilson line proceeds via a single off-shell gluon. These contributions can be evaluated by contracting the tree-level soft current for single soft-gluon emission with the one-loop triple-gluon vertex with one leg off-shell. The one-particle irreducible contributions to the latter can be found for example in Ref.~\cite{Davydychev:1996pb}, while propagator-type diagrams are textbook material.

Finally, Fig.~\ref{fig:main1line} contains the remaining non-vanishing diagrams with two virtual gluons coupling to the Wilson line.

The colour structure of single Wilson-line emissions can be decomposed in the following basis:
\begin{gather} \label{eq:ColorBasis}
    f^{a_1 b d} f^{a_2 c d} \comm{\mathbf{T}_i^{b}}{\mathbf{T}_i^c} = \frac{C_A}{2} i f^{a_1a_2c} \mathbf{T}_i^{c} \; , \qquad
    f^{a_1 b d} f^{a_2 c d} \acomm{\mathbf{T}_i^{b}}{\mathbf{T}_i^c} \; .
\end{gather}
These two structures can be rewritten in terms of just one upon using colour conservation:
\begin{equation}
\begin{aligned} \label{eq:1WilsonTo2Wilson}
    f^{a_1 b d} f^{a_2 c d} \sum_i \comm{\mathbf{T}_i^{b}}{\mathbf{T}_i^c} F_{\alpha_1\alpha_2}(q_1,q_2,p_i) = - f^{a_1 b d} f^{a_2 c d} \sum_{i \neq j} \mathbf{T}_i^{b} \mathbf{T}_j^{c} \big( F_{\alpha_1\alpha_2}(q_1,q_2,p_i) - F_{\alpha_1\alpha_2}(q_1,q_2,p_j) \big) \; , \\[.2cm]
    f^{a_1 b d} f^{a_2 c d} \sum_i \acomm{\mathbf{T}_i^{b}}{\mathbf{T}_i^c} G_{\alpha_1\alpha_2}(q_1,q_2,p_i) = - f^{a_1 b d} f^{a_2 c d} \sum_{i \neq j} \mathbf{T}_i^{b} \mathbf{T}_j^{c} \big( G_{\alpha_1\alpha_2}(q_1,q_2,p_i) + G_{\alpha_1\alpha_2}(q_1,q_2,p_j) \big) \; ,
\end{aligned}
\end{equation}
where $F_{\alpha_1\alpha_2}(q_1,q_2,p_i)$ and $G_{\alpha_1\alpha_2}(q_1,q_2,p_i)$ represent the functional dependence of single Wilson-line diagrams after stripping their colour structures. The use of Eqs.~\eqref{eq:1WilsonTo2Wilson} is necessary in order to verify the Ward identity for $\Delta \mathbf{J}^{(1)}{}^{a_1 a_2}_{\alpha_1 \alpha_2}(q_1, q_2)$
, since it is not satisfied by the diagrams of Figs.~\ref{fig:1eikonal1line} and \ref{fig:main1line} alone.

\begin{figure}
\begin{tabular}{cccc}
    \includegraphics[scale=0.7, page=34]{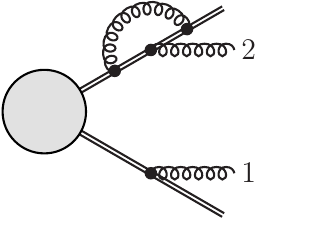}&    
    \includegraphics[scale=0.7, page=35]{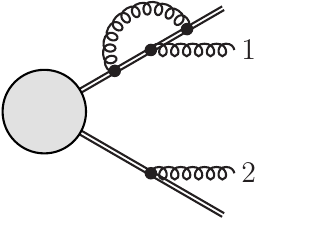}&    
    \includegraphics[scale=0.7, page=36]{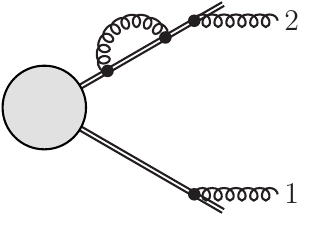}&    
    \includegraphics[scale=0.7, page=37]{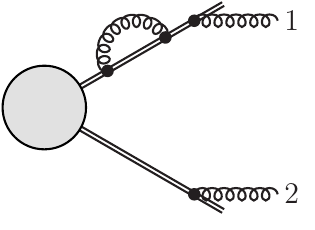}\\
    \includegraphics[scale=0.7, page=42]{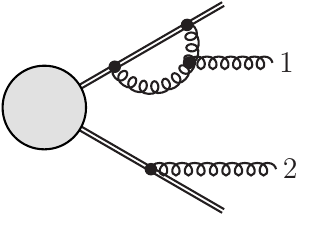}&    
    \includegraphics[scale=0.7, page=43]{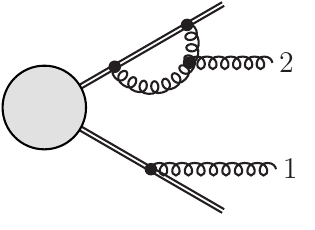}&
    \includegraphics[scale=0.7, page=44]{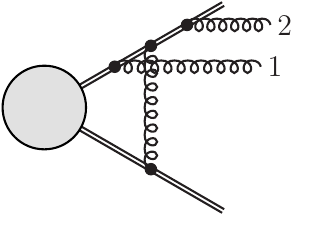}&    
    \includegraphics[scale=0.7, page=45]{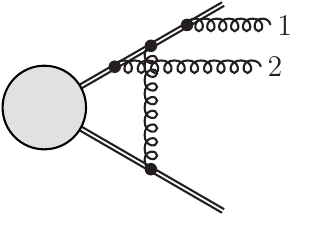}\\     
    \includegraphics[scale=0.7, page=46]{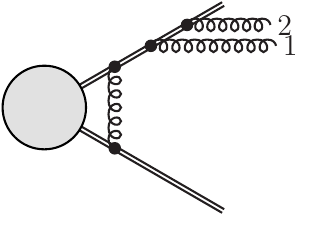}&     
    \includegraphics[scale=0.7, page=47]{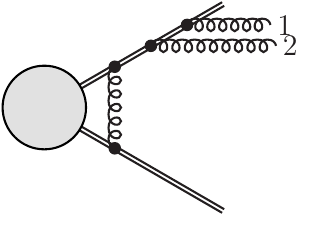}&     
    \includegraphics[scale=0.7, page=48]{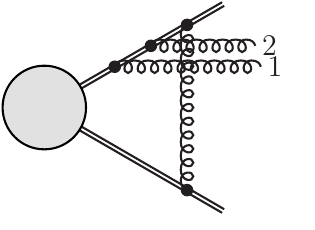}&     
    \includegraphics[scale=0.7, page=49]{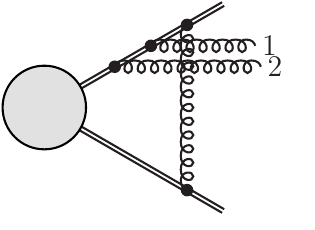}\\     
    \includegraphics[scale=0.7, page=50]{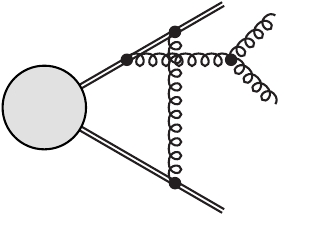}&     
    \includegraphics[scale=0.7, page=51]{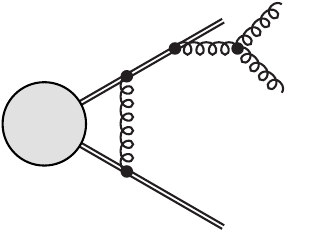}& 
\end{tabular}
    \caption{Scaleless diagrams with two Wilson lines. \label{fig:scaleless2lines}}
\end{figure}

\begin{figure}
\begin{tabular}{cccc}
    \includegraphics[scale=0.7, page=38]{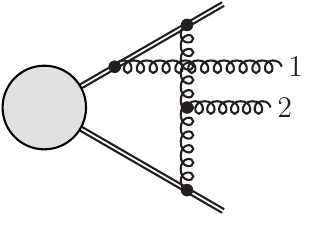}&    
    \includegraphics[scale=0.7, page=39]{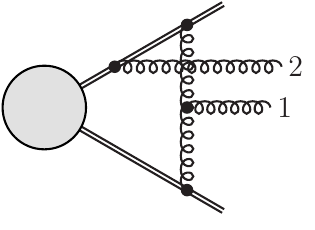}&    
    \includegraphics[scale=0.7, page=40]{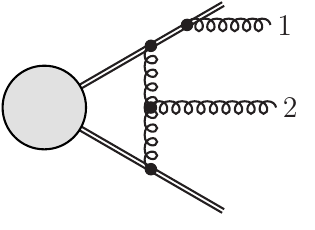}&    
    \includegraphics[scale=0.7, page=41]{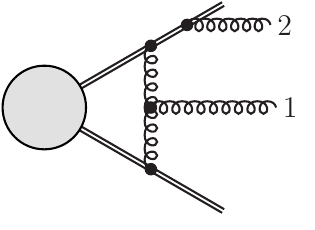}\\  
\end{tabular}
    \caption{Diagrams with two Wilson lines that contribute to $\mathbf{J}^{(1)} \mathbf{J}^{(0)}$ and $\Delta \mathbf{J}^{(1)}$. \label{fig:iterated2lines}}
\end{figure}

\begin{figure}
\begin{tabular}{cccc}
    \includegraphics[scale=0.7, page=52]{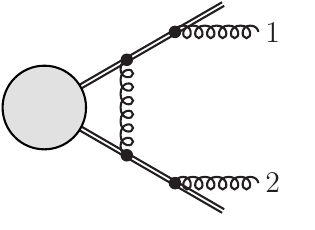}&    
    \includegraphics[scale=0.7, page=53]{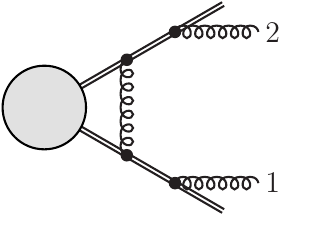}&    
    \includegraphics[scale=0.7, page=58]{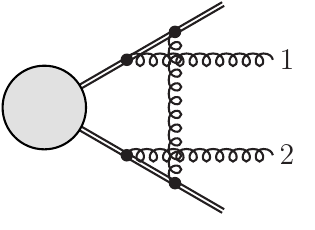}&    
    \includegraphics[scale=0.7, page=59]{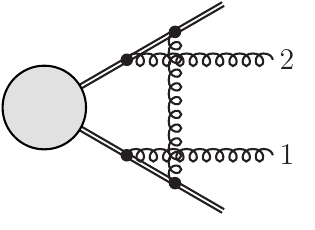}\\   
    \includegraphics[scale=0.7, page=54]{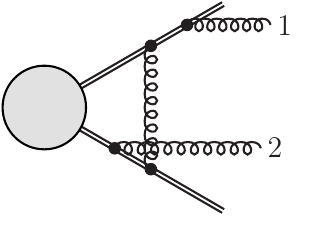}& 
    \includegraphics[scale=0.7, page=55]{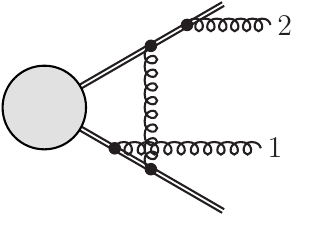}&    
    \includegraphics[scale=0.7, page=56]{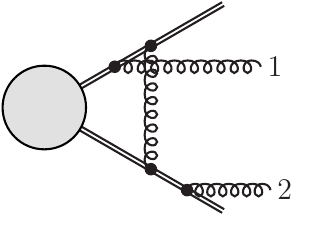}&    
    \includegraphics[scale=0.7, page=57]{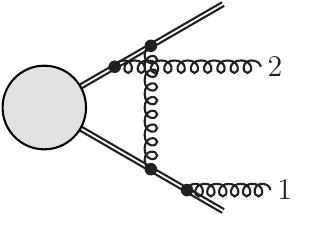}\\  
    \includegraphics[scale=0.7, page=60]{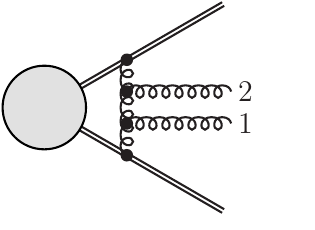}&  
    \includegraphics[scale=0.7, page=61]{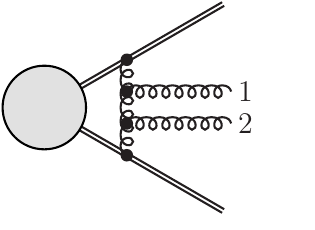}&    
    \includegraphics[scale=0.7, page=62]{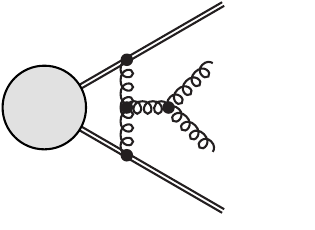}&    
    \includegraphics[scale=0.7, page=63]{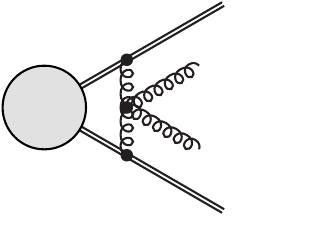}\\   
\end{tabular}
    \caption{Remaining diagrams with two Wilson lines that only contribute to $\Delta \mathbf{J}^{(1)}$.}
    \label{fig:main2lines}
\end{figure}

Diagrams involving two different Wilson lines corresponding to partons $i$ and $j$ are depicted in Figs.~\ref{fig:scaleless2lines}, \ref{fig:iterated2lines} and \ref{fig:main2lines}. The presence of two hard momenta, $p_i$ and $p_j$, means that the respective Feynman integrals are non-vanishing when they are either proportional to the factor \eqref{eq:prefactor1} or to:
\begin{equation}
    \bigg( \frac{\mu^2 p_i \cdot p_j}{(q_k \cdot p_i)(q_l \cdot p_j)} \bigg)^{\epsilon} \; , \qquad k,l \in \{1,2\} \; .
\end{equation}
From this and the arguments presented above in the case of a single Wilson line follows that all diagrams in Fig.~\ref{fig:scaleless2lines} are scaleless and do not contribute. Indeed, the occurring Feynman integrals only depend on $q_1 \cdot p_i$ and/or $q_2 \cdot p_i$, but neither on $q_1 \cdot p_i$ together with $q_2 \cdot p_j$, nor on $q_1 \cdot q_2$.

Further insight into the remaining diagrams is gained by considering the eikonal identity in case of two emissions from the same Wilson line. For definiteness, let us assume that the Wilson line carries an outgoing momentum $p_i$, and one of the emitted gluons has momentum $q_1$ and colour $a_1$ (soft gluon), while the other is real or virtual and has momentum $k$ and colour $c$. The gluons may be emitted in two different orderings, hence the corresponding Feynman diagrams may be summed and the respective Feynman integrand will contain the following term:
\begin{multline} \label{eq:eikonalIdentity}
    \frac{\mathbf{T}_i^{a_1}}{p_i \cdot q_1} \frac{\mathbf{T}_i^{c}}{p_i \cdot (q_1 + k)} + \frac{\mathbf{T}_i^{c}}{p_i \cdot k} \frac{\mathbf{T}_i^{a_1}}{p_i \cdot (q_1 + k)} = \frac{1}{2} \acomm{\frac{\mathbf{T}_i^{a_1}}{p_i \cdot q_1}}{\frac{\mathbf{T}_i^{c}}{p_i \cdot k}} - \frac{p_i \cdot (q_1 - k)}{2p_i \cdot (q_1 + k)} \, \comm{\frac{\mathbf{T}_i^{a_1}}{p_i \cdot q_1}}{\frac{\mathbf{T}_i^{c}}{p_i \cdot k}} \; .
\end{multline}
Applying this relation to the diagrams in Fig.~\ref{fig:iterated2lines} yields two contributions. The first of these is just:
\begin{equation}
\begin{split}
\frac{1}{2} \, \Big( \acomm{\mathbf{J}^{(1)}{}^{a_1}_{\alpha_1}(q_1)}{ \mathbf{J}^{(0)}{}^{a_2}_{\alpha_2}(q_2)} &+ \acomm{\mathbf{J}^{(1)}{}^{a_2}_{\alpha_2}(q_2)}{ \mathbf{J}^{(0)}{}^{a_1}_{\alpha_1}(q_1)} \Big) = \\[.2cm] &\Big( \mathbf{J}^{(1)}{}^{a_1}_{\alpha_1}(q_1) \mathbf{J}^{(0)}{}^{a_2}_{\alpha_2}(q_2) + \mathbf{J}^{(1)}{}^{a_2}_{\alpha_2}(q_2) \mathbf{J}^{(0)}{}^{a_1}_{\alpha_1}(q_1) \Big) \\[.2cm] &- \frac{1}{2} \, \Big( \comm{\mathbf{J}^{(1)}{}^{a_1}_{\alpha_1}(q_1)}{ \mathbf{J}^{(0)}{}^{a_2}_{\alpha_2}(q_2)} + \comm{\mathbf{J}^{(1)}{}^{a_2}_{\alpha_2}(q_2)}{ \mathbf{J}^{(0)}{}^{a_1}_{\alpha_1}(q_1)} \Big) \; .
\end{split}
\end{equation}
The bracket on the second line matches the respective term in Eq.~\eqref{eq:DeltaJ1def}, while the bracket on the third line contributes to $\Delta \mathbf{J}^{(1)}{}^{a_1a_2}_{\alpha_1\alpha_2}(q_1,q_2)$ with the colour structure:
\begin{equation} \label{eq:color2lines}
    f^{a_1bd} f^{a_2cd} \mathbf{T}_i^{b} \mathbf{T}_j^{c} \; , \qquad i \neq j \; .
\end{equation}
The second contribution resulting from the application of the eikonal identity \eqref{eq:eikonalIdentity} to the diagrams in Fig.~\ref{fig:iterated2lines} is proportional to exactly the same colour structure \eqref{eq:color2lines}.

Turning to the last set of diagrams depicted in Fig.~\ref{fig:main2lines}, we notice that the anti-commutator in the eikonal identity \eqref{eq:eikonalIdentity} now only yields scaleless integrals, hence only the commutator actually contributes to $\Delta \mathbf{J}^{(1)}{}^{a_1a_2}_{\alpha_1\alpha_2}(q_1,q_2)$. In consequence, the only colour structure present in contributions due to two Wilson lines is that given in \eqref{eq:color2lines}.

In view of the above colour-structure analysis, the soft current has the following form after using colour conservation:
\begin{empheq}[box=\fbox]{equation} \label{eq:ColorDecomposition}
    \Delta \mathbf{J}^{(1)}{}^{a_1 a_2}_{\alpha_1 \alpha_2} = \frac{1}{q_1 \cdot q_2} \, f^{a_1 b d} f^{a_2 c d} \sum_{i \neq j} \mathbf{T}_i^{b} \mathbf{T}_j^c \, J_{\alpha_1 \alpha_2}(q_1,q_2,p_i,p_j) \; ,
\end{empheq}
where $J_{\alpha_1 \alpha_2}(q_1,q_2,p_i,p_j)$ is a dimensionless rank-two tensor. Due to the symmetry properties of the colour structure and the symmetry of the soft current w.r.t.\ the exchange of the two soft gluons, there is:
\begin{equation}
    J_{\alpha_1 \alpha_2}(q_1,q_2,p_i,p_j) = J_{\alpha_2 \alpha_1}(q_2,q_1,p_j,p_i) \; .
\end{equation}
Furthermore, the Ward identity is satisfied trivially, since colour conservation has already been exploited by using Eqs.~\eqref{eq:1WilsonTo2Wilson}:
\begin{equation}
    q_1^{\alpha_1} J_{\alpha_1 \alpha_2}(q_1,q_2,p_i,p_j) = q_2^{\alpha_2} J_{\alpha_1 \alpha_2}(q_1,q_2,p_i,p_j) = 0 \; .
\end{equation}
According to Eq.~\eqref{eq:asymptotics}, the soft current describes amplitude asymptotics after contraction with polarisation vectors. The Ward identity makes it possible to absorb any dependence of $J_{\alpha_1 \alpha_2}(q_1,q_2,p_i,p_j)$ on $q_1^{\alpha_2}$ and $q_2^{\alpha_1}$ into the polarisation vectors by making them orthogonal to both $q_1$ and $q_2$. Assuming general polarisation vectors, $\epsilon_j \equiv \epsilon(q_j)$, $j=1,2$, we therefore define:
\begin{equation}
    \epsilon'_1{}^{\alpha_1} = \epsilon_1^{\beta_1} \Big( \delta_{\beta_1}^{\alpha_1} - \frac{q_{2\beta_1} q_1^{\alpha_1}}{q_1 \cdot q_2} \Big) \; , \qquad
    \epsilon'_2{}^{\alpha_2} = \epsilon_2^{\beta_2} \Big( \delta_{\beta_2}^{\alpha_2} - \frac{q_{1\beta_2} q_2^{\alpha_2}}{q_1 \cdot q_2} \Big) \; , \qquad
    q_{1,2} \cdot \epsilon'_{1,2} = 0 \; .
\end{equation}
The contraction of the polarisation vectors with the soft current is now given by:
\begin{equation}
    \epsilon_1^{\alpha_1 *} \epsilon_2^{\alpha_2 *} J_{\alpha_1 \alpha_2}(q_1,q_2,p_i,p_j) = \epsilon'_1{}^{\alpha_1 *} \epsilon'_2{}^{\alpha_2 *} J_{\alpha_1 \alpha_2}(q_1,q_2,p_i,p_j) \; .
\end{equation}
The modified polarisation vectors project onto the transverse space w.r.t.\ both $q_1$ and $q_2$. The spin structure of $J_{\alpha_1 \alpha_2}(q_1,q_2,p_i,p_j)$ is therefore given by the five tensors:
\begin{equation}
    g_{\perp\alpha_1\alpha_2} \; , \qquad p_{i\perp\alpha_1}p_{i\perp\alpha_2} \; , \qquad p_{j\perp\alpha_1}p_{j\perp\alpha_2}  \; , \qquad p_{i\perp\alpha_1}p_{j\perp\alpha_2} \; , \qquad p_{j\perp\alpha_1}p_{i\perp\alpha_2} \; ,
\end{equation}
where:
\begin{align}
    &g_{\perp\alpha_1\alpha_2} \equiv g_{\alpha_1\alpha_2} - \frac{q_{1\alpha_1}q_{2\alpha_2} + q_{2\alpha_1}q_{1\alpha_2}}{q_1 \cdot q_2} \; , \qquad &q_{1,2}^{\alpha_1} \, g_{\perp\alpha_1\alpha_2} = g_{\perp\alpha_1\alpha_2} \, q_{1,2}^{\alpha_2} = 0 \; , \\[.2cm]
    &p_{i,j\perp} \equiv p_{i,j} - \Big( \frac{p_{i,j} \cdot q_2}{q_1 \cdot q_2} \Big) \, q_1 - \Big( \frac{p_{i,j} \cdot q_1}{q_2 \cdot q_1} \Big) \, q_2 \; , \qquad &p_{i,j\perp} \cdot q_{1,2} = 0 \; .
\end{align}
The four vectors $q_1$, $q_2$, $p_i$ and $p_j$ define a four-dimensional Minkowski space as long as the kinematic configuration is non-singular, i.e.\ no two vectors are collinear. We therefore split the $d$-dimensional metric tensor into an $\epsilon$-dimensional, $\tilde{g}$, and a four-dimensional, $\hat{g}$, component:
\begin{gather}
    g_{\alpha_1\alpha_2} \equiv \tilde{g}_{\alpha_1\alpha_2} + \hat{g}_{\alpha_1\alpha_2} \; , \\[.2cm]
    \hat{g}_{\alpha_1\alpha_2} \equiv \sum_{ij=1}^4 \big( G^{-1} \big)_{ij} k_{i\alpha_1} k_{j\alpha_2} \; , \qquad
    G_{ij} \equiv k_i \cdot k_j \; , \qquad
    ( k_i )_{i=1}^4 \equiv ( q_1,q_2,p_i,p_j ) \; .
\end{gather}
This implies in particular:
\begin{equation}
    g_{\perp\alpha_1\alpha_2} = \tilde{g}_{\alpha_1\alpha_2} + \hat{g}_{\perp\alpha_1\alpha_2} \; ,
\end{equation}
where $\hat{g}_{\perp\alpha_1\alpha_2}$ can be decomposed in terms of $p_{i,j\perp}$:
\begin{equation} \label{eq:gperpDecomposition}
    \hat{g}_{\perp\alpha_1\alpha_2} = \frac{1}{\sin^2\big( \sphericalangle(\bm{p}_{iT}, \bm{p}_{jT}) \big)} \bigg[ \frac{p_{i\perp\alpha_1} p_{i\perp\alpha_2}}{p_{i\perp}^2} + \frac{p_{j\perp\alpha_1} p_{j\perp\alpha_2}}{p_{j\perp}^2} - \frac{( p_{i\perp} \cdot p_{j\perp} )^2}{p_{i\perp}^2 p_{j\perp}^2} \, \frac{p_{i\perp\alpha_1} p_{j\perp\alpha_2} + p_{j\perp\alpha_1} p_{i\perp\alpha_2}}{p_{i\perp} \cdot p_{j\perp}} \bigg] \; ,
\end{equation}
with $\bm{p}_{i,jT}$ the three-vector components of $p_{i,j\perp}$ in the rest frame of $q_{1,2}$, and:
\begin{gather}
    \sin^2\big( \sphericalangle(\bm{p}_{iT}, \bm{p}_{jT}) \big) = 1 - \frac{( p_{i\perp} \cdot p_{j\perp} )^2}{p_{i\perp}^2 p_{j\perp}^2} = -\frac{\det(G)}{4 (p_i \cdot q_1) (p_i \cdot q_2) (p_j \cdot q_1) (p_j \cdot q_2)} \; , \\[.2cm]
    p_{i\perp} \cdot p_{j\perp} = - \frac{(p_i \cdot q_1)(p_j \cdot q_2) + (p_i \cdot q_2) (p_j \cdot q_1) - (p_i \cdot p_j) (q_1 \cdot q_2)}{q_1 \cdot q_2} \; , \qquad
    p_{i,j\perp}^2 = - \frac{2(p_{i,j} \cdot q_1)(p_{i,j} \cdot q_2)}{q_1 \cdot q_2} \; .
\end{gather}
We use this decomposition to simplify our expressions as described Section~\ref{sec:ExactResults}. In consequence, however, the soft current depends separately on $\tilde{g}$ and $\hat{g}_{\perp}$ rather than on $g_{\perp}$ only.

In four dimensions, the two polarisation vectors of the soft gluons can only assume one of four helicity configurations: $++$, $--$, $+-$ and $-+$. Due to the symmetry of QCD w.r.t.\ space inversion, the soft current for configurations $--$ and $-+$ can be recovered from its values for $++$ and $+-$ configurations respectively. As already pointed out, aside from the soft-current contribution proportional to $\tilde{g}$, $J_{\alpha_1 \alpha_2}(q_1,q_2,p_i,p_j)$ is effectively four-dimensional. Hence, it is desirable to decompose it according to the independent helicity configurations. This is achieved as follows:
\begin{empheq}[box=\fbox]{equation} \label{eq:TensorDecomposition}
\begin{split}
J&{}_{\alpha_1 \alpha_2}(q_1, q_2, p_i, p_j) \equiv \tilde{J}(q_1, q_2, p_i, p_j) \big[ \tilde{g}_{\alpha_1 \alpha_2} \big] + \hat{J}_{++} (q_1, q_2, p_i, p_j) \big[ \hat{g}_{\perp\alpha_1 \alpha_2} \big] \\[.2cm]
 &+J_{++}(q_1, q_2, p_i, p_j) \bigg[  \frac{q_1 \cdot q_2}{(p_i \cdot q_1) (p_j \cdot q_2)} \big( p_{i\perp\alpha_1} p_{j\perp\alpha_2} - p_{j\perp\alpha_1} p_{i\perp\alpha_2} \big) \bigg] \\[.2cm] 
 &+J_{+-}(q_1, q_2, p_i, p_j) \bigg[ \hat{g}_{\perp \alpha_1 \alpha_2} - \frac{2 \, p_{i\perp\alpha_1} p_{i\perp\alpha_2}}{p_{i\perp}^2} \bigg] + J_{+-}(q_2, q_1, p_j, p_i) \bigg[ \hat{g}_{\perp \alpha_1 \alpha_2} - \frac{2 \, p_{j\perp\alpha_1} p_{j\perp\alpha_2}}{p_{j\perp}^2} \bigg] \; .
 \end{split}
\end{empheq}

The tensor coefficients of $\hat{J}_{++}$ and $J_{++}$ vanish for the helicity configurations $+-$ and $-+$, while those of $J_{+-}$ vanish for the helicity configurations $++$ and $--$. We prove this statement as follows. Consider the polarisation vectors $\epsilon'_{1\pm}$ and $\epsilon'_{2\pm}$ in the rest frame of the two soft gluons. Assume the standard relation between polarisation vectors for opposite helicities, $\epsilon^{\prime\,*}_{1,2\pm} = \epsilon^{\prime}_{1,2\mp}$, as well as the standard normalisation $\epsilon^{\prime}_{1,2\lambda} \cdot \epsilon^{\prime\,*}_{1,2\lambda'} = -\delta_{\lambda\lambda'}$. Since both polarisation vectors are orthogonal to $q_1$ and $q_2$, $\epsilon_{\pm} \equiv \epsilon'_{1\pm}$ constitute a basis for $\epsilon'_{2\pm}$. By definition of helicity, the polarisation vectors are eigenstates of rotations around the direction of motion of the respective gluons. Since a clockwise rotation around $\bm{q}_1$ is a counter-clockwise rotation around $\bm{q}_2$, there is:
\begin{equation}
    \epsilon'_{2+} = e^{i \phi} \epsilon_{-} \; ,  \qquad
    \epsilon^{\prime \, \alpha_1 *}_{1+} \epsilon^{\prime \, \alpha_2 *}_{2-} = e^{i\phi} \, \epsilon^{\alpha_1}_{-} \epsilon^{\alpha_2}_{-} \; , \qquad
    \epsilon^{\prime \, \alpha_1 *}_{1+} \epsilon^{\prime \, \alpha_2 *}_{2+} = e^{-i\phi} \, \epsilon^{\alpha_1}_{-} \epsilon^{\alpha_2}_{+} \; ,
\end{equation}
for some phase $\phi$. In consequence, the tensor coefficients of $\hat{J}_{++}$ and $J_{++}$ indeed vanish for the $+-$ and $-+$ helicity configurations:
\begin{equation}
\begin{gathered}
    \epsilon^{\prime \, \alpha_1 *}_{1+} \epsilon^{\prime \, \alpha_2 *}_{2-} \, \big[ \hat{g}_{\perp\alpha_1\alpha_2} \big] = e^{i \phi} \big( \epsilon_- \cdot \epsilon_- \big) = 0 \; , \\[.2cm]
    \epsilon^{\prime \, \alpha_1 *}_{1+} \epsilon^{\prime \, \alpha_2 *}_{2-} \, \big[ p_{i\perp\alpha_1} p_{j\perp\alpha_2} - p_{j\perp\alpha_1} p_{i\perp\alpha_2} \big] = e^{i \phi} \big( \big( \epsilon_- \cdot p_i \big) \big( \epsilon_- \cdot p_j \big) - \big( \epsilon_- \cdot p_j \big) \big( \epsilon_- \cdot p_i \big) \big) = 0 \; .
\end{gathered}
\end{equation}
Since the transverse components of the momentum four-vectors can be decomposed in the $\epsilon_\pm$ basis:
\begin{equation}
    p_{i,j\perp} = - \big( \epsilon_+ \cdot p_{i,j} \big) \, \epsilon_- - \big( \epsilon_- \cdot p_{i,j} \big) \, \epsilon_+ \; , \qquad p_{i,j\perp}^2 = -2 \, (p_{i,j} \cdot \epsilon_+)(p_{i,j} \cdot \epsilon_-) \; ,
\end{equation}
there is:
\begin{gather}
    \epsilon^{\prime \, \alpha_1 *}_{1+} \epsilon^{\prime \, \alpha_2 *}_{2+} \, \bigg[ \frac{2 \, p_{i,j\perp\alpha_1} p_{i,j\perp\alpha_2}}{p_{i,j\perp}^2} \bigg] = e^{-i\phi} \, \frac{2 \, (p_{i,j} \cdot \epsilon_-)(p_{i,j} \cdot \epsilon_+)}{p_{i,j\perp}^2} = - e^{-i\phi} \; ,
\end{gather}
and independently:
\begin{gather}
    \epsilon^{\prime \, \alpha_1 *}_{1+} \epsilon^{\prime \, \alpha_2 *}_{2+} \, \big[ \hat{g}_{\alpha_1\alpha_2} \big] = e^{-i\phi} \, \epsilon_- \cdot \epsilon_+ = - e^{-i\phi} \; .
\end{gather}
This proves the vanishing of the tensor coefficients of $J_{+-}$ for the helicity configurations $++$ and $--$.

It is worth noticing that $\hat{g}_{\perp\alpha_1\alpha_2}$ is a scalar in the transverse space, while $p_{i\perp\alpha_1} p_{j\perp\alpha_2} - p_{j\perp \alpha_1} p_{i\perp \alpha_2}$ is a longitudinal vector along $q_{1,2}$. On the other hand, the rank-two tensors in the square brackets defining $J_{+-}$ in Eq.~\eqref{eq:TensorDecomposition} are symmetric and traceless in the transverse space, and thus correspond to an irreducible spin two representation of rotations.

The tree-level soft current \eqref{eq:2gluonJ0} can also be rewritten in a form similar to Eqs.~\eqref{eq:DeltaJ1def}, \eqref{eq:ColorDecomposition} and \eqref{eq:TensorDecomposition}:
\begin{equation} \label{eq:2gluonJ0alt}
    \begin{split}
        \bm{\mathcal{J}}^{(0)}&{}^{a_1 a_2}_{\alpha_1 \alpha_2}(q_1, q_2) = \frac{1}{2} \left( \bm{\mathcal{J}}^{(0)}{}^{a_1}_{\alpha_1}(q_1) \mathbf{J}^{(0)}{}^{a_2}_{\alpha_2}(q_2) + \bm{\mathcal{J}}^{(0)}{}^{a_2}_{\alpha_2}(q_2) \mathbf{J}^{(0)}{}^{a_1}_{\alpha_1}(q_1) \right) \\[.2cm]
        &\begin{split} + \frac{1}{C_A} \, f^{a_1 b d} f^{a_2 c d} \sum_{i \neq j} \mathbf{T}_i^{b} \mathbf{T}_j^c \bigg\{
        &2 g_{\perp\alpha_1 \alpha_2} \frac{(p_i \cdot q_1)(p_j \cdot q_2) - (p_i \cdot q_2)(p_j \cdot q_1)}{(q_1 \cdot q_2) \big( p_i \cdot (q_1+q_2) \big) \big( p_j \cdot (q_1+q_2) \big)} \\[.2cm]
        &+ \frac{p_{i\perp\alpha_1} p_{j\perp\alpha_2} - p_{j\perp\alpha_1} p_{i\perp\alpha_2}}{(p_i \cdot q_1)(p_j \cdot q_2)} + \frac{p_{i\perp\alpha_1} p_{j\perp\alpha_2} + p_{j\perp\alpha_1} p_{i\perp\alpha_2}}{(p_i \cdot q_1)(p_j \cdot q_2)} \\[.2cm]
        &- \frac{2p_{i\perp\alpha_1} p_{i\perp\alpha_2}}{(p_i \cdot q_1) \big( p_i \cdot (q_1+q_2) \big)} -\frac{2p_{j\perp\alpha_1} p_{j\perp\alpha_2}}{(p_j \cdot q_2) \big( p_j \cdot (q_1+q_2) \big)} \bigg\} \; .
        \end{split}
    \end{split}
\end{equation}
In obtaining this result, we have used Eqs.~\eqref{eq:1gluonJ0alt}, \eqref{eq:ColorBasis} and \eqref{eq:1WilsonTo2Wilson}. We have also dropped terms that vanish by colour conservation. Notice, however, that we have not applied the decomposition of the metric tensor Eq.~\eqref{eq:gperpDecomposition}. Eq.~\eqref{eq:2gluonJ0alt} is useful in verifying the infrared and ultraviolet singularities of $\bm{\mathrm{J}}^{(1)}{}^{a_1 a_2}_{\alpha_1 \alpha_2}(q_1, q_2)$ with the help of Eq.\ (20) in Ref.~\cite{Catani:2021kcy}. In particular, the leading singularity of $J_{\alpha_1 \alpha_2}(q_1, q_2, p_i, p_j)$ is $-2 ( q_1 \cdot q_2) /\epsilon^2$ multiplied with the expression in the curly bracket on the r.h.s.\ of Eq.~\eqref{eq:2gluonJ0alt}. Our results presented in Section~\ref{sec:ExpandedResults} confirm this value, as well as the value of the subleading singularity. The agreement is, nevertheless, only obtained after application of Eq.~\eqref{eq:gperpDecomposition}.

\subsection{Exact dependence on the spacetime dimension} \label{sec:ExactResults}

In order to obtain the actual values of the functions $\hat{J}_{++}$, $J_{++}$, $J_{+-}$ and $\tilde{J}$ defined through Eqs.~\eqref{eq:DeltaJ1def}, \eqref{eq:ColorDecomposition} and \eqref{eq:TensorDecomposition}, we have performed a Passarino-Veltman tensor reduction \cite{Passarino:1978jh} of the Feynman integrals corresponding to all diagrams, followed by an integration-by-parts reduction using the software package \textsc{Kira} \cite{Maierhofer:2017gsa, Klappert:2020nbg}. The final results are expressed in terms of one-loop master integrals defined as follows:
\begin{equation}
  I^{(d)}_{a_1a_2a_3a_4a_5}\equiv \mu^{2\epsilon} \int \frac{\textrm{d}^dl}{i\pi^{d/2}}
  \frac{1}{[l^2]^{a_1}[(l+q_1)^2]^{a_2}[(l-q_2)^2]^{a_3}[p_i\cdot(l+q_1)]^{a_4}[-p_j\cdot(l-q_2)]^{a_4}} \; .
  \label{eq:MI}
\end{equation}
The integral $I^{(d)}_{11111}$ has been replaced by $I^{(d+2)}_{11111}$ using the methods of Refs.~\cite{Tarasov:1996br, Tarasov:1997kx}, see Eq.~\eqref{eq:DimShift} in Appendix~\ref{app:Integrals}. This has the advantage of removing spurious singularities in $\epsilon$. Furthermore, it demonstrates that the soft current expanded up to $\order{\epsilon^0}$ does not require the knowledge of the complicated integral $I^{(d+2)}_{11111}$. This latter property requires the use of the decomposition  \eqref{eq:gperpDecomposition} of the metric tensor. The values of the integrals are given in Appendix~\ref{app:Integrals}. As a test, we have verified that the soft-current satisfies the Ward identity \eqref{eq:Ward} without substituting the integrals\footnote{Of course, the test is only non-trivial if the transverse projection is not applied.}. 

Our results exact in the dimension of spacetime $d$ are as follows:
\begin{flalign*}
  \hat{J}_{++}(q_1, q_2, p_i,p_j) &= J_{++}(q_1, q_2, p_i, p_j) \frac{s_{1i} s_{2j} - s_{2i} s_{1j} + s_{12} s_{ij}}{s_{1i} s_{2j}} + \epsilon I_{11111}^{(d + 2)} \, \frac{1}{2} s_{12} s_{ij}  && \\ 
  & \quad +\epsilon I_{01100}^{(d)} \, \frac{2 }{3 - 2 \epsilon} \left(1 - \frac{2}{1 - \epsilon} \frac{ n_l T_F}{C_A} \right) \left(\frac{s_{1i}}{s_{1i} + s_{2i}} - \frac{s_{1j}}{s_{1j} + s_{2j}} \right)    \; , && \stepcounter{equation}\tag{\theequation}
\end{flalign*}
\begin{flalign*}
  J_{++}(q_1, q_2, p_i, p_j) &= - I_{11110}^{(d)} \, \frac{s_{12} s_{1i}}{4} - I_{11101}^{(d)} \, \frac{s_{12} s_{2j}}{4} - I_{01111}^{(d)} \, \frac{(s_{1i} + s_{2i} ) (s_{1j} + s_{2j}) - s_{12} s_{ij}}{8} && \\  
  & \quad - \epsilon I_{11111}^{(d + 2)} \, \frac{1}{4} (s_{1i} s_{2j}-s_{2i} s_{1j} + s_{12}  s_{ij})  \; , && \stepcounter{equation}\tag{\theequation}
\end{flalign*}
\begin{flalign*}
  &J_{+-}(q_1, q_2, p_i,p_j)= I_{01100}^{(d)} \left(\frac{2}{\epsilon} - 4 \right) \frac{s_{1i}}{s_{1i} + s_{2i}} \left[1 +  \frac{(s_{1i} + s_{2i}) s_{2j}}{2s_{12}(p_{i \perp} \cdot p_{j \perp})} \right] + I_{10011}^{(d)} \, \frac{ s_{ij}^2}{8 p_{i \perp} \cdot p_{j \perp}} && \\ 
  & \quad + \left(I_{11101}^{(d)} \, 2 s_{12} s_{2j}  + I_{11011}^{(d)} \, s_{1i} (s_{1j} + s_{2j})  \right) \frac{ s_{1i} s_{2i} (s_{1j}^2 - s_{2j}^2) }{8 s_{12} (p_{i \perp} \cdot p_{j \perp}) \left( (s_{1i} + s_{2i}) (s_{1j} + s_{2j} ) - s_{12} s_{ij} \right) } && \\ \stepcounter{equation}\tag{\theequation}
  &  \quad - \left( I_{11110}^{(d)} \, 2 s_{12} s_{1i}  + I_{10111}^{(d)} \, s_{2j} (s_{1i} + s_{2i} )  \right) \bigg [ \frac{s_{1i} s_{2i} (s_{1j}^2 - s_{2j}^2) }{8 s_{12} (p_{i \perp} \cdot p_{j \perp}) \left( (s_{1i} + s_{2i}) (s_{1j} + s_{2j} ) - s_{12} s_{ij} \right) } && \\
  & \hspace{9.5cm} +  \frac{s_{1i} s_{2j}- s_{2i} s_{1j}}{8 s_{12} (p_{i \perp} \cdot p_{j \perp})} + \frac{s_{1i} - s_{2i} }{4 (s_{1i} + s_{2i})} \bigg ] && \\ 
  & \quad + \epsilon I_{11111}^{(d + 2)}\, \frac{s_{1i} s_{2i}}{4 s_{12} (p_{i \perp} \cdot p_{j \perp})}  \left[ (s_{1j} - s_{2j})^2 - \frac{(s_{1j}^2 - s_{2j}^2) (s_{1i} s_{1j} - s_{2i} s_{2j})}{(s_{1i} + s_{2i}) (s_{1j} + s_{2j}) - s_{12} s_{ij}} \right] \; ,
\end{flalign*}
\begin{flalign*}
  &\tilde{J}(q_1,q_2, p_i, p_j)= I_{01100}^{(d)}\, \frac{2}{3 - 2 \epsilon} \left(\frac{(1 - \epsilon) (3 - 5 \epsilon)}{\epsilon} - \frac{2 \epsilon}{(1 - \epsilon)} \frac{n_l T_F }{C_A} \right)  \frac{s_{1i} s_{2j} - s_{2i} s_{1j}}{(s_{1i} + s_{2i}) (s_{1j} + s_{2j})} && \\ 
  & \quad - \left(  I_{11101}^{(d)} \, 2 s_{12} s_{2j} + I_{11011}^{(d)} \, s_{1i} (s_{1j} + s_{2j})  \right) \frac{1}{4} \bigg [ \frac{s_{1i} s_{1j} - s_{2i} s_{2j}}{(s_{1i} + s_{2i}) (s_{1j} + s_{2j}) - s_{12} s_{ij}} - \frac{s_{1j} - s_{2j}}{s_{1j} + s_{2j}} \bigg ] && \\ 
  & \quad + \left(I_{11110}^{(d)} \, 2 s_{12} s_{1i}  + I_{10111}^{(d)} \, s_{2j} (s_{1i} + s_{2i})   \right)\frac{1}{4} \bigg [ \frac{s_{1i} s_{1j} - s_{2i} s_{2j}}{(s_{1i} + s_{2i}) (s_{1j} + s_{2j}) - s_{12} s_{ij}} - \frac{s_{1i} - s_{2i}}{s_{1i} + s_{2i}} \bigg ] && \stepcounter{equation}\tag{\theequation} \\ 
  & \quad - I_{11111}^{(d + 2)} \, \frac{1}{2} \bigg [s_{12} s_{ij} + \epsilon \left( (s_{1i} - s_{2i}) (s_{1j} - s_{2j}) + s_{12} s_{ij} - \frac{(s_{1i} s_{1j} - s_{2i} s_{2j} )^2}{(s_{1i} + s_{2i}) (s_{1j} + s_{2j}) - s_{12} s_{ij}} \right) \bigg ] \; ,
\end{flalign*}
where
\begin{equation}
    s_{12} \equiv 2 q_1 \cdot q_2 \; , \qquad 
    s_{1,2\,i,j} \equiv 2 q_{1,2} \cdot p_{i,j} \; , \qquad 
    s_{ij} \equiv 2 p_i \cdot p_j \; .
\end{equation}
The functions $\tilde{J}$ and $\hat{J}_{++}$ depend on the number of massless quarks $n_l$. 

\subsection{\texorpdfstring{Expansion accurate at $\order{\epsilon^0}$}{Expansion accurate at O(E0)}} \label{sec:ExpandedResults}

The results of the previous section can be expanded in $\epsilon$ after substitution of the master integrals. An expansion up to $\order{\epsilon^0}$ is sufficient to describe the double-soft limit of the finite remainders of one-loop amplitudes defined in the `t Hooft-Veltman scheme, i.e.\ with four-dimensional external momenta and polarisations. The function $\tilde{J}$, although singular at $\epsilon = 0$, does not contribute in this case, since the tensor $\tilde{g}_{\alpha_1\alpha_2}$ is $\epsilon$-dimensional. The truncated expansions are: 
\begin{flalign*}
  \hat{J}_{++} (q_1, q_2, p_i, p_j) &= J_{++}(q_1, q_2, p_i, p_j) \frac{s_{1i} s_{2j} - s_{2i} s_{1j} + s_{12} s_{ij}}{s_{1i} s_{2j}} && \\
  &\quad + \left(\frac{2}{3} - \frac{4}{3C_A} n_l T_F \right) \left( \frac{s_{1i}}{s_{1i} + s_{2i}} - \frac{s_{1j}}{s_{1j} + s_{2j}} \right)
  + \mathcal{O}(\epsilon) \stepcounter{equation}\tag{\theequation} \label{eq:gluonJppHat} \; , &&
\end{flalign*}
\begin{flalign*}
  &J_{++}(q_1, q_2, p_i, p_j) = r_\Gamma  \left(\frac{-s_{12} - i0^+}{\mu^2} \right)^{-\epsilon}
  \bigg \{-\frac{2}{\epsilon^2} - \frac{1}{\epsilon}\ln \! \left(\frac{s_{12} s_{ij}}{s_{1i} s_{2j}} \right)  && \\
  \stepcounter{equation}\tag{\theequation} \label{eq:gluonJpp}
  & \quad - \frac{1}{2} \left[\ln^2 \! \left( \frac{s_{12} s_{ij}}{(s_{1i} + s_{2i})(s_{1j} + s_{2j})} \right) + \ln^2 \! \left(\frac{s_{1i}}{s_{1i} + s_{2i}}\right)  + \ln^2 \! \left( \frac{s_{2j}}{s_{1j} + s_{2j}} \right) \right] && \\
  & \quad - \left[\text{Li}_2 \! \left(1 - \frac{s_{12} s_{ij}}{(s_{1i} + s_{2i}) (s_{1j} + s_{2j})} \right) + \text{Li}_2 \! \left(1 - \frac{s_{1i}}{s_{1i} + s_{2i}} \right) + \text{Li}_2 \! \left(1 - \frac{s_{2j}}{s_{1j} + s_{2j}} \right) \right]  + \mathcal{O}(\epsilon) \bigg \} \; , &&
\end{flalign*}
\begin{flalign*}
  &J_{+-}(q_1, q_2, p_i, p_j) = r_\Gamma  \left(\frac{-s_{12} - i0^+}{\mu^2} \right)^{-\epsilon} &&\\
  &\quad\times\bigg\{\left(- \frac{2}{\epsilon^2} - \frac{1}{\epsilon} 
  \ln \! \left(\frac{ s_{12}s_{ij}}{s_{1i} s_{2j}} \right)
  -\frac{1}{2}\ln^2 \! \left( \frac{ s_{12}s_{ij}}{s_{1i} s_{2j}} \right) 
  -\frac{\pi^2}{3} \right)\left(\frac{-2s_{2i}}{s_{1i}+s_{2i}} -
  \frac{s_{2i}s_{1j}}{s_{12}(p_{i \perp} \cdot p_{j \perp})}\right) &&\\
  &\quad + \ln \! \left( \frac{s_{2j}}{s_{1j} + s_{2j}} \right) 
  \ln \! \left(\frac{ s_{12}s_{ij}}{s_{1i} (s_{1j} + s_{2j})} \right)
  \frac{s_{1i} s_{2i} (s_{1j}^2 - s_{2j}^2) }{ s_{12} (p_{i \perp} \cdot p_{j \perp})
  \left( (s_{1i} + s_{2i}) (s_{1j} + s_{2j} ) - s_{12} s_{ij} \right) } \stepcounter{equation}\tag{\theequation} \label{eq:gluonJpm} && \\
  &\quad - \ln \! \left(\frac{s_{1i}}{s_{1i} + s_{2i}} \right)
  \ln \! \left(\frac{ s_{12}s_{ij}}{s_{2j} (s_{1i} + s_{2i})} \right) \bigg [
  \frac{s_{1i} s_{2i} (s_{1j}^2 - s_{2j}^2) }{ s_{12} (p_{i \perp} \cdot p_{j \perp})
  \left( (s_{1i} + s_{2i}) (s_{1j} + s_{2j} ) - s_{12} s_{ij} \right) } && \\
  & \hspace{9cm} +  \frac{s_{1i} s_{2j}- s_{2i} s_{1j}}{s_{12} (p_{i \perp} \cdot p_{j \perp})} + 
  2\,\frac{s_{1i} - s_{2i} }{s_{1i} + s_{2i}} \bigg ] + \mathcal{O}(\epsilon) \bigg \} \; . &&
\end{flalign*}
The above results are valid for both partons $i$ and $j$ outgoing. Analytic continuation to other configurations is straightforward and described in Appendix~\ref{app:Integrals}.

By contracting Eq.~\eqref{eq:TensorDecomposition} with appropriate polarisation vectors, we have verified with the help of the software package \textsc{S@M} \cite{Maitre:2005uu} that the above expressions agree with the results of Ref.~\cite{Zhu:2020ftr} with the exception of a sign difference in the term proportional to the colour structure\footnote{We have replaced this colour structure using Eqs.~\eqref{eq:ColorBasis} and \eqref{eq:1WilsonTo2Wilson}, while it has been kept as is in Ref.~\cite{Zhu:2020ftr}.} $if^{a_1a_2c} \bm{\mathrm{T}}^c_i$ for the $++$ helicity configuration. Performing further tests, in particular those described in Appendix~\ref{app:Tests}, we have convinced ourselves that the sign we have obtained is correct. Notice that the term in question is not singular and its value cannot be verified using the general structure of singularities of the soft current given in Eq.\ (20) of Ref.~\cite{Catani:2021kcy}. On the other hand, this term is generated by the diagrams of Fig.~\ref{fig:1eikonal1line} and can thus be evaluated using known results from the literature as described in Section~\ref{sec:Diagrams}. This completely avoids the methods listed in Section~\ref{sec:ExactResults} and thus constitutes an additional test of correctness.

\section{Soft quark-anti-quark pair emission} \label{sec:SoftQuarks}

\begin{figure}
\begin{center}
\begin{tabular}{cccc}
    \includegraphics[page=1, scale=0.7]{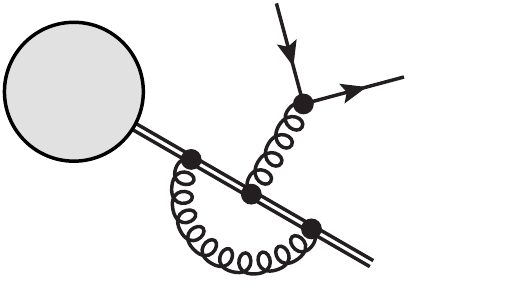}
    \includegraphics[page=2, scale=0.7]{diagrams/diagrams-qq-crop.pdf}
\end{tabular}
    \caption{Diagrams for soft quark-anti-quark pair emission with one Wilson line that correspond to scaleless integrals. \label{fig:scaleless1line-quarks}}
\end{center}
\begin{center}
\begin{tabular}{cccc}
    \includegraphics[page=3, scale=0.7]{diagrams/diagrams-qq-crop.pdf}&
    \includegraphics[page=4, scale=0.7]{diagrams/diagrams-qq-crop.pdf}&
    \includegraphics[page=5, scale=0.7]{diagrams/diagrams-qq-crop.pdf}&
    \includegraphics[page=6, scale=0.7]{diagrams/diagrams-qq-crop.pdf}\\
    \includegraphics[page=7, scale=0.7]{diagrams/diagrams-qq-crop.pdf}&
\end{tabular}
    \caption{Diagrams with only one gluon coupling to the Wilson line. \label{fig:1gluon1line-quarks}}
\end{center}
\begin{center}
\begin {tabular}{cccc}
    \includegraphics[page=8, scale=0.7]{diagrams/diagrams-qq-crop.pdf}&
    \includegraphics[page=9, scale=0.7]{diagrams/diagrams-qq-crop.pdf}&
    \includegraphics[page=10, scale=0.7]{diagrams/diagrams-qq-crop.pdf}&
\end{tabular}
    \caption{Remaining diagrams with only one Wilson line. \label{fig:main1line-quarks}}
\end{center}
\begin{center}
\begin{tabular}{cccc}
    \includegraphics[page=11, scale=0.7]{diagrams/diagrams-qq-crop.pdf}&
    \includegraphics[page=12, scale=0.7]{diagrams/diagrams-qq-crop.pdf}&
\end{tabular}
    \caption{Diagrams with two Wilson lines that correspond to scaleless integrals. \label{fig:scaless2lines-quarks}}
\end{center}
\begin{center}
\begin{tabular}{cccc}
    \includegraphics[page=13, scale=0.7]{diagrams/diagrams-qq-crop.pdf}&
    \includegraphics[page=14, scale=0.7]{diagrams/diagrams-qq-crop.pdf}&
    \includegraphics[page=15, scale=0.7]{diagrams/diagrams-qq-crop.pdf}&
\end{tabular}
    \caption{Remaining diagrams with two Wilson lines. \label{fig:main2lines-quarks}}
\end{center}
\end{figure}

The case of a soft quark-anti-quark pair can be analysed with the same methods as those used for two soft gluons. The occurring diagrams are depicted in Figs.~\ref{fig:scaleless1line-quarks}, \ref{fig:1gluon1line-quarks}, \ref{fig:main1line-quarks}, \ref{fig:scaless2lines-quarks} and \ref{fig:main2lines-quarks}. In particular, the diagrams in Figs.~\ref{fig:scaleless1line-quarks} and \ref{fig:scaless2lines-quarks} do not contribute as they generate scaleless Feynman integrals by the same argument as the one provided in Section~\ref{sec:Diagrams}. Fig.~\ref{fig:1gluon1line-quarks} is an off-shell one-loop gluon-quark vertex attached to the Wilson line. The most complex contributions are to be found in Figs.~\ref{fig:main1line-quarks} and \ref{fig:main2lines-quarks}.

Since a single soft quark does not generate a singularity in a scattering amplitude, the soft current for quark-anti-quark pair emission does not contain iterated contributions in contrast to Eq.~\eqref{eq:DeltaJ1def}. On the other hand, it can be defined by analogy to Eq.~\eqref{eq:asymptotics} and expanded as in Eq.~\eqref{eq:Jexp}. The decomposition of the one-loop contribution analogous to Eqs.~\eqref{eq:ColorDecomposition} and \eqref{eq:TensorDecomposition} for gluons is:
\begin{empheq}[box=\fbox]{multline} \label{eq:J1quarks}
  \mathbf{J}^{(1)}_{a_1 a_2}(q_1,\lambda_1,q_2,\lambda_2) =  \frac{1}{q_1 \cdot q_2} \sum_{i\neq j} (T^bT^c)_{a_1a_2} \mathbf{T}^b_i \mathbf{T}^c_j \, \\[.2cm] 
  \times \left[ \bar{u}(q_1,\lambda_1) \left( \frac{\slashed{p}_i}{p_i \cdot (q_1 + q_2)} \, J_{q\bar{q}}(q_1,q_2, p_i, p_j)  - \frac{\slashed{p}_j}{p_j \cdot (q_1 + q_2)} \, J_{q\bar{q}}(q_2,q_1, p_j, p_i) \right) v(q_2,\lambda_2) \right] \; ,
\end{empheq}
where we have kept the outgoing quark bi-spinor, $\bar{u}(q_1,\lambda_1)$, and the outgoing anti-quark bi-spinor, $v(q_2,\lambda_2)$, in the soft current at variance with the gluon case where polarisation vectors have been separted from the soft current. By chirality conservation of massless QCD, the only helicity configurations of the quark-anti-quark pair that yield non-vanishing contributions are $+-$ and $-+$. For comparison, the tree-level soft current can be obtained from Eq.~\eqref{eq:1gluonJ0} by attaching a tree-level gluon-quark vertex:
\begin{equation} \label{eq:J0quarks}
    \mathbf{J}^{(0)}_{a_1 a_2} = - \sum_{i} T^c_{a_1a_2} \mathbf{T}^c_i \, \frac{\bar{u}(q_1,\lambda_1) \, \slashed{p}_i \, v(q_2,\lambda_2)}{2 ( q_1 \cdot q_2 ) \big( p_i \cdot (q_1  + q_2) \big)} \; .
\end{equation}

The colour structure in Eq.~\eqref{eq:J1quarks} is determined as follows. Diagrams of Figs.~\ref{fig:1gluon1line-quarks} and \ref{fig:main1line-quarks} can be expressed in terms of:
\begin{equation}
    (T^bT^c)_{a_1a_2} \comm{\mathbf{T}^b_i}{\mathbf{T}^c_i} = - \frac{C_A}{2} \, T^c_{a_1a_2} \mathbf{T}^c_i \; , \qquad
    (T^bT^c)_{a_1a_2} \acomm{\mathbf{T}^b_i}{\mathbf{T}^c_i} \; .
\end{equation}
Using colour conservation, both can be reduced to:
\begin{equation} \label{eq:ColorStructureQuarks}
    (T^bT^c)_{a_1a_2} \mathbf{T}^b_i \mathbf{T}^c_j \; , \qquad i \neq j \; ,
\end{equation}
with the same method as in the case of gluons, see Eqs.~\eqref{eq:1WilsonTo2Wilson}. The colour structure \eqref{eq:ColorStructureQuarks} is also the only one needed for the diagrams of Fig.~\ref{fig:main2lines-quarks}. Furthermore, up to colour conservation, the tree-level soft current \eqref{eq:J0quarks} can be rewritten as follows:
\begin{equation}
    \bm{\mathcal{J}}^{(0)}_{a_1 a_2} = -\frac{1}{C_A} \frac{1}{q_1 \cdot q_2} \sum_{i\neq j} (T^bT^c)_{a_1a_2} \mathbf{T}^b_i \mathbf{T}^c_j \, \left[ \bar{u}(q_1,\lambda_1) \left( \frac{\slashed{p}_i}{p_i \cdot (q_1  + q_2)} - \frac{\slashed{p}_j}{p_j \cdot (q_1  + q_2)} \right) v(q_2,\lambda_2) \right] \; .
\end{equation}
As in the case of the gluon soft current, this expression can be used to verify the singularities of $\mathbf{J}^{(1)}_{a_1 a_2}$ using Eq.\ (20) of Ref.~\cite{Catani:2021kcy}. For instance, the leading singularity is just $-2C_F/\epsilon^2 \bm{\mathcal{J}}^{(0)}_{a_1 a_2}$.

The spin structure in Eq.~\eqref{eq:J1quarks} is a straightforward consequence of chirality conservation in massless QCD. Indeed, after use of Dirac algebra for contracted indices, the soft current can only contain terms of the form:
\begin{equation}
    \bar{u}(q_1,\lambda_1) \, \slashed{v}_1 \cdots \slashed{v}_n \, v(q_2,\lambda_2) \; , \qquad v_i \in \{ q_1, q_2, p_i, p_j \}  \; , \qquad \text{$n$ - odd} \; .
\end{equation}
By equations of motion and Dirac algebra, these can be reduced to:
\begin{equation}
    \bar{u}(q_1,\lambda_1) \, \slashed{p}_i \, v(q_2,\lambda_2) \; , \qquad
    \bar{u}(q_1,\lambda_1) \, \slashed{p}_j \, v(q_2,\lambda_2) \; .
\end{equation}
The relative sign between the two structures in Eq.~\eqref{eq:J1quarks} follows from charge-conjugation invariance of QCD. Denoting the charge conjugation operator with $\mathbb{C}$, there is:
\begin{align}
    &\mathbb{C} \, A^\mu_{ab} \, \mathbb{C}^{-1} = - A^{\mu\,\ast}_{ab} \; , \qquad  \; A^\mu_{ab} \equiv A^{\mu\, c} T^c_{ab} \; , \\[.2cm]
    &\mathbb{C} \ket{q_1a_1\lambda_1,q_2a_2\lambda_2} = \ket{q_2a_2\lambda_2,q_1a_1\lambda_1} \; , \label{eq:ConQQbar}
\end{align}
where $\ket{qa\lambda,q'a'\lambda'}$ is a quark-anti-quark state with $q,a,\lambda$ the quantum numbers of the quark, and $q',a',\lambda'$ those of the anti-quark. Eq.~\eqref{eq:ConQQbar} is a consequence of the anti-commutation of the quark and anti-quark creation operators, and of the sign difference between the intrinsic charge parity of the quark and the anti-quark. This sign difference is equivalent to $\bar{u}(q_1,\lambda_1) \, \gamma^\mu \, v(q_2,\lambda_2) = - \bar{u}(q_2,\lambda_2) \, \gamma^\mu \, v(q_1,\lambda_1)$. On the other hand, charge conjugation of the Wilson-line operator $W_i$ is achieved with the replacement $A^c_\mu \bm{\mathrm{T}}_i^c \, \to \, \mathbb{C} \, A^c_\mu \, \mathbb{C}^{-1} \bm{\mathrm{T}}_i^c$, which implies the replacement of colour-charge operators $T^c \bm{\mathrm{T}}_i^c \to - (T^c)^T \bm{\mathrm{T}}_i^c$. Applying these relations to the r.h.s.\ of Eq.~\eqref{eq:JbyWilson} for a quark-anti-quark final state yields:
\begin{equation} \label{eq:ChargeConjugation}
    \mathbf{J}^{(1)}_{a_1 a_2}(q_1,\lambda_1,q_2,\lambda_2) = \mathbf{J}^{(1)}_{a_2 a_1}(q_2,\lambda_2,q_1,\lambda_1) 
    \bigg|_{T^c \bm{\mathrm{T}}_i^c \to - (T^c)^T \bm{\mathrm{T}}_i^c} \; .
\end{equation}
The structure of Eq.~\eqref{eq:J1quarks} is a direct consequence.

Our result for the $J_{q\bar{q}}$ function with exact dependence on the dimension $d$ of spacetime reads:
\begin{flalign*}
  J_{q\bar{q}}&(q_1,q_2, p_i, p_j) = \\
  &- I^{(d)}_{01100} \left[\frac{3}{\epsilon}+\frac{1}{6-4\epsilon}-\frac{5}{2} + \epsilon - \left(\frac{2}{\epsilon}-1+2\epsilon\right)\frac{C_F}{C_A}
  -\frac{4(1-\epsilon)}{3-2\epsilon}\frac{n_lT_F}{C_A}\right] &&\\
  &+ \frac{1}{8}\frac{s_{1i} + s_{2i}}{(s_{1i}+s_{2i}) (s_{1j}+s_{2j})-s_{12} s_{ij}} \bigg[
  I^{(d)}_{11011} \, s_{1i}(s_{1j}+s_{2j})^2 + I^{(d)}_{11101} \, 2 s_{12}s_{2j} (s_{1j} + s_{2j})\stepcounter{equation}\tag{\theequation}&&\\
  &\hspace{4cm}+\frac{I^{(d)}_{10111} \, (s_{1i}+s_{2i})s_{2j} +I^{(d)}_{11110} \, 2s_{12}s_{1i}}{ (s_{1i}+s_{2i})} ((s_{1i}+s_{2i})(s_{1j}+s_{2j})-2s_{12}s_{ij})&&\\
  &\hspace{4cm}-\epsilon I^{(d+2)}_{11111} \, 2 \left(  (s_{1j}+s_{2j}) (s_{1i} s_{2j}-s_{1j} s_{2i})+ s_{12}s_{ij} (s_{1j}-s_{2j}) \right) \bigg] \; .
\end{flalign*}
Upon expansion in $\epsilon$ up to $\order{\epsilon^0}$, we further obtain:
\begin{flalign*}
  J_{q\bar{q}}&(q_1,q_2, p_i, p_j) = r_\Gamma \left(\frac{-s_{12} - i0^+}{\mu^2} \right)^{-\epsilon} &&\\
  &\times\bigg\{\frac{2C_F}{C_A}\frac{1}{\epsilon^2} + \left(\frac{3C_F}{C_A}+\frac{4n_lT_F}{3C_A}-\frac{11}{3}
  +\ln \! \left(\frac{s_{12}s_{ij}}{s_{1i}s_{2j}}\right)\right)\frac{1}{\epsilon}&&\\
  &\hspace{4cm}
  + \frac{8C_F}{C_A}+\frac{20n_lT_F}{9C_A}-\frac{76}{9}
  +\frac{1}{2}\ln^2 \! \left(\frac{s_{12}s_{ij}}{s_{1i}s_{2j}}\right)+\frac{\pi^2}{3} \stepcounter{equation}\tag{\theequation} \label{eq:quarkJpm} &&\\
  &\quad+\frac{(s_{1i} + s_{2i}) s_{12}}{(s_{1i}+s_{2i})(s_{1j}+s_{2j})-s_{12}s_{ij}} \bigg[\ln \! \left(\frac{s_{1i}}{s_{1i}+s_{2i}}\right)
  \ln \! \left(\frac{s_{12}s_{ij}}{s_{2j}(s_{1i}+s_{2i})}\right) \left(\frac{s_{1j}+s_{2j}}{s_{12}}-\frac{2s_{ij}}{s_{1i}+s_{2i}}\right) &&\\
  &\hspace{6cm}+ \ln \! \left( \frac{s_{2j}}{s_{1j} + s_{2j}} \right) \ln \! \left(\frac{ s_{12}s_{ij}}{s_{1i} (s_{1j} + s_{2j})} \right)
  \frac{s_{1j}+s_{2j}}{s_{12}}\bigg] + \mathcal{O}(\epsilon)\bigg\} \; ,
\end{flalign*}
which agrees with Ref.~\cite{Catani:2021kcy} and, up to a global sign (see Appendix~\ref{app:Erratum}), with Ref.~\cite{Zhu:2020ftr}. The above expansion is valid for both partons $i$ and $j$ outgoing. Analytic continuation to other configurations is straightforward and described in Appendix~\ref{app:Integrals}.

\section{Outlook}

With this publication, we provide the soft current for the emission of either two gluons or a quark-anti-quark pair as an exact expression in Conventional Dimensional Regularisation (CDR). Real and virtual momenta, gluon polarisation vectors and the virtual gluon field are all $d$-dimensional in CDR. In some applications, however, it might be desirable to use a variant of dimensional regularisation, for example the four-dimensional helicity scheme \cite{Bern:1991aq}. A translation between schemes can be achieved without a recalculation of the soft current. The appropriate formula is given in Eq.~(31) of Ref.~\cite{Catani:2021kcy}. As far as cross section evaluation is concerned, variants of dimensional regularisation are only used in the calculation of the phase-space integrals of the exact one-loop amplitude regulated with a subtraction term. In this case, therefore, only the $\order{\epsilon^0}$ expansion of the soft current is required with non-CDR regularisation. In consequence, the results for the soft current with explicit dependence on the variant of dimensional regularisation provided in Refs.~\cite{Zhu:2020ftr, Catani:2021kcy} are sufficient.

As alluded to in the introduction, our $d$-dimensional results are necessary for the evaluation of phase-space integrals of the soft-subtraction term. A detailed recent discussion explaining the reasoning and providing a method for determining the order of the necessary expansion in $\epsilon$ can be found in Section 2.3 of Ref.~\cite{Czakon:2022fqi}. According to that reference, if the kinematic configuration is non-singular apart from the non-hierarchical double-soft limit, then the soft current is only needed to $\order{\epsilon}$. One case, where higher order expansion terms become indispensable, is due to the presence of an iterated soft limit. This case can be treated by combining two single soft-gluon emission currents as has been discussed in Section 3.2 of Ref.~\cite{Zhu:2020ftr}. On the other hand, the result can also be obtained directly from our formulae, because $I_{11111}^{(d+2)}$ does not contribute in the iterated soft limit. Another case, where an $\order{\epsilon^2}$ expansion of the soft current is needed, corresponds to the triple-collinear double-soft limit. This limit can be derived from our results, but it is more efficient to use the expressions provided in Ref.~\cite{Czakon:2022fqi}. For N3LO applications, the expansion of the soft current to $\order{\epsilon^3}$ and $\order{\epsilon^4}$ is only needed in presence of additional soft and collinear limits, which can be treated with formulae derived for NLO and NNLO applications in Refs.~\cite{Campbell:1997hg, Catani:1998nv, Bern:1998sc, Kosower:1999rx, Bern:1999ry, Catani:1999ss, Catani:2000pi, Czakon:2011ve, Bierenbaum:2011gg, Czakon:2018iev, Catani:2011st, Sborlini:2013jba}. These formulae are exact in $d$ and do not suffer from any shortcomings in consequence.

In conclusion, there are two limitation of our work. If the results were to be used as part of an N$n$LO, $n > 3$ cross section calculation, then the integral $I_{11111}^{(d+2)}$ would have to be calculated to higher order in $\epsilon$, but no other improvements would be needed. On the other hand, an N3LO calculation of a cross section involving massive partons, as for example typical in applications to top-quark physics, requires the soft current for $p_i^2 \neq 0$. Comparison of the massless \cite{Catani:2000pi} and massive \cite{Bierenbaum:2011gg,Czakon:2018iev} single soft-gluon current demonstrates that one should expect a substantially more complicated double-emission soft current in the massive case than the results provided here. We leave this problem to future work.

We note, finally, that in order to avoid mistakes in using our results, the functions presented in Sections~\ref{sec:ExactResults}, \ref{sec:ExpandedResults} and \ref{sec:SoftQuarks} are available in electronic format (\textsc{Mathematica}) together with the preprint version of this work.

\begin{acknowledgments}
This work was supported by the Deutsche Forschungsgemeinschaft (DFG) under grant 396021762 - TRR 257: Particle Physics Phenomenology after the Higgs Discovery. The Feynman diagrams were drawn using \textsc{Axodraw}~\cite{Vermaseren:1994je,Collins:2016aya}.
\end{acknowledgments}

\appendix

\section{Integrals} \label{app:Integrals}

With the exception of $I_{11111}^{(d)}$, the master integrals corresponding to various sequences of propagator powers, $a_i \in \{0,1\}$, in Eq.~\eqref{eq:MI} can be obtained with exact $\epsilon$-dependence by introducing Feynman parameters. The calculation is simplified by applying the Cheng-Wu theorem and only retaining the Feynman parameters of the quadratic propagators in the $\delta$-function constraint of the Feynman parameter integral. The master integrals have been previously evaluated in Ref.~\cite{Zhu:2020ftr}. We agree with these results up to an overall factor in the case of $I^{(d)}_{11011}$, see Appendix~\ref{app:Erratum}. As a further test, we have compared numerical values of the integrals for random external momenta against the results obtained with the software package \textsc{pySecDec} \cite{Borowka:2017idc} up to $\mathcal{O}(\epsilon)$. Our results for the master integrals are as follows:
\begin{flalign}
I_{01100}^{(d)} = r_\Gamma \left[\frac{-s_{12} - i 0^+}{\mu^2} \right]^{-\epsilon} \frac{1}{\epsilon (1 - 2 \epsilon)} \; , &&
\end{flalign}
\begin{flalign}
I_{11110}^{(d)} = r_\Gamma \left[ \frac{-s_{12} - i 0^+}{\mu^2} \right]^{-\epsilon} \frac{1}{\epsilon^2} \frac{4}{s_{12} s_{1i}} {}_2F_1 \! \left(1, - \epsilon, 1 - \epsilon; - \frac{s_{2i}}{s_{1i}} \right) \; , &&
\end{flalign}
\begin{flalign}
I_{11101}^{(d)} = I_{11110}^{(d)} \, \Big\vert_{ q_1 \leftrightarrow q_2, \, p_i \leftrightarrow p_j } \; , &&
\end{flalign}
\begin{flalign}
I_{10011}^{(d)} =  r_\Gamma  \left[ \frac{-s_{12} - i0^+}{\mu^2} \right]^{-\epsilon} \frac{\Gamma(1 + \epsilon) \Gamma(1 - 2 \epsilon)}{\epsilon^2 \Gamma(1 - \epsilon)} \frac{4}{s_{ij}} \left( \frac{(s_{1i} + i0^+) (s_{2j} + i0^+)}{(s_{12} + i 0^+)(s_{ij} + i 0^+)} \right)^{-\epsilon} \; , &&
\end{flalign}
\begin{flalign*}
I_{11011}^{(d)}  = r_\Gamma  \left[ \frac{-s_{12} - i0^+}{\mu^2} \right]^{-\epsilon} \frac{\Gamma(1 + \epsilon)  \Gamma (1 - 2 \epsilon)}{\epsilon^2 \Gamma(1 - \epsilon)} &  \frac{4}{s_{1i} s_{2j}} \left( \frac{(s_{1i} + i 0^+) (s_{2j} + i 0^+)}{(s_{12} + i0^+)(s_{ij} + i 0^+)} \right)^{-\epsilon} && \stepcounter{equation}\tag{\theequation}\\ 
& \qquad \qquad \times {}_2F_1 \! \left(1 + \epsilon, 1, 1 - \epsilon; - \frac{s_{1j}}{s_{2j}} \right) \; , &&
\end{flalign*}
\begin{flalign}
I_{10111}^{(d)} = I_{11011}^{(d)} \, \Big \vert_{ q_1 \leftrightarrow q_2, \, p_i \leftrightarrow p_j } \; , &&
\end{flalign}
\begin{flalign*}
I_{01111}^{(d)} &=-r_\Gamma \left[\frac{-s_{12} - i 0^+}{\mu^2} \right]^{-\epsilon} \frac{1}{\epsilon (1 + \epsilon)} \frac{8}{s_{12} s_{ij}}  \left(\frac{(s_{1i} + s_{2i} + i 0^+)(s_{1j} + s_{2j} + i 0^+)}{( s_{12} + i 0^+)(s_{ij} + i 0^+)} \right)^{-1 -\epsilon} && \stepcounter{equation}\tag{\theequation}
\\ 
& \hspace{4cm} \times {}_2F_1 \! \left( 1 + \epsilon, 1 + \epsilon, 2 + \epsilon; 1 - \frac{(s_{12} + i 0^+)(s_{ij} + i 0^+) }{(s_{1i} + s_{2i} + i 0^+)(s_{1j} + s_{2j} + i 0^+)} \right) \; ,
\end{flalign*} 
where:
\begin{equation}
    r_\Gamma \equiv \frac{\Gamma^2(1-\epsilon)\Gamma(1+\epsilon)}{\Gamma(1-2\epsilon)} \; ,\qquad
    s_{12} \equiv 2 q_1 \cdot q_2 \; , \qquad 
    s_{1,2\,i,j} \equiv 2 q_{1,2} \cdot p_{i,j} \; , \qquad 
    s_{ij} \equiv 2 p_i \cdot p_j \; ,
\end{equation}
for outgoing momenta $p_i$ and $p_j$. The causal $+i0^+$ has been included where necessary to allow for analytic continuation due to crossing of the partons $i$ and/or $j$ to the initial state, which amounts to changing the sign of the respective momentum. Details are discussed below. Finally, the hypergeometric functions ${}_2F_1$ can be expanded with the help of the software package \textsc{HypExp} \cite{Maitre:2005uu, Maitre:2007kp, Huber:2005yg}.

At present, we are unable to evaluate the integral $I_{11111}^{(d)}$ with exact $\epsilon$ dependence. The result is thus given as a Laurent expansion. For reasons explained in Section~\ref{sec:ExactResults}, this integral has been replaced by $I_{11111}^{(d + 2)}$ using the dimensional shift relation:
\begin{flalign*} \label{eq:DimShift}
&I^{(d)}_{11111} = - \frac{ s_{1i} s_{2j} - s_{2i} s_{1j}+ s_{12} s_{ij}}{2 s_{1i} s_{2j} s_{12}} I_{01111}^{(d)} + \frac{I_{11101}^{(d)}}{s_{1i}} + \frac{I_{11110}^{(d)}}{s_{2j}} && \\ 
& \quad + \frac{s_{1i} - s_{2i}}{s_{12}} \frac{I_{10111}^{(d)}}{2s_{1i}} - \frac{s_{1j} - s_{2j}}{s_{12}} \frac{I_{11011}^{(d)}}{2 s_{2j}} + \frac{(s_{1i} - s_{2i})(s_{1j} - s_{2j}) + s_{12} s_{ij}}{s_{1i} s_{2j} s_{12}} \, \epsilon I_{11111}^{(d + 2)} && \\ \stepcounter{equation}\tag{\theequation}
& \quad + \frac{s_{1i} s_{1j} - s_{2i} s_{2j}}{(s_{1i} + s_{2i})(s_{1j} + s_{2j}) - s_{12} s_{ij}} \bigg [ \frac{I_{11101}^{(d)}}{s_{1i}} - \frac{I_{11110}^{(d)}}{s_{2j}}  - \frac{s_{1i} + s_{2i}}{s_{12}} \frac{I_{10111}^{(d)}}{2 s_{1i}} + \frac{s_{1j} + s_{2j}}{ s_{12}} \frac{I_{11011}^{(d)}}{2 s_{2j}} && \\ 
& \hspace{11.5cm} - \frac{s_{1i} s_{1j} - s_{2i} s_{2j}}{s_{1i} s_{2j} s_{12}} \, \epsilon I_{11111}^{(d + 2)}  \bigg ] \; , && \\ 
\end{flalign*}
which has been obtained using the methods of Refs.~\cite{Tarasov:1996br, Tarasov:1997kx}. The main advantage of the integral replacement is that $I_{11111}^{(d + 2)}$ is finite, which can be understood both from the perspective of soft power counting as well as from the well-known fact that there are no infrared divergences in six and higher dimensions. Clearly, the knowledge of $I_{11111}^{(d + 2)}$ to $\order{\epsilon^n}$ is sufficient to obtain $I^{(d)}_{11111}$ to $\order{\epsilon^{n+1}}$. In particular, the result for $I_{11111}^{(d)}$ to $\order{\epsilon^0}$ can be obtained without the value of $I_{11111}^{(d + 2)}$. This is important, since this integral has been previously evaluated in Ref.~\cite{Zhu:2020ftr} by the simplified-differential-equation method \cite{Papadopoulos:2014lla}. We agree with Ref.~\cite{Zhu:2020ftr} up to obvious typos, see Appendix~\ref{app:Erratum}.

The $(d + 2)$-dimensional integral can be evaluated using the software package \textsc{PolyLogTools} \cite{Duhr:2019tlz} by direct integration of the Feynman parameter representation. Interestingly, $I_{11111}^{(d + 2)}$ has already been evaluated in Ref.~\cite{Czakon:2022fqi} as the double-soft limit of an integral occurring in the calculation of one-loop triple-collinear splitting functions. Despite this fact, we have tested the result for $I_{11111}^{(d)}$ against numerical values provided by \textsc{pySecDec} up to $\mathcal{O}(\epsilon^2)$.

The result for $I_{11111}^{(d + 2)}$ is available from Ref.~\cite{Czakon:2022fqi} in electronic form in terms of multiple polylogarithms. Here, we reproduce the value of the integral at $\epsilon = 0$ expressed in terms of ordinary logarithms, di-logarithms and tri-logarithms. The expansion to $\order{\epsilon}$ is attached in electronic form to the preprint of the present publication as well:
\begin{flalign*}
-(r_1 &- r_2) x_{1} y_{1} \frac{(s_{1i} + s_{2i}) (s_{1j} + s_{2j})}{4} I_{11111}^{(d + 2)} = \text{Li}_3 \! \left(\frac{r_{1} x_{1} y_{1}}{u_{3}}\right)-\text{Li}_2 \! \left(\frac{r_{1} x_{1} y_{1}}{u_{3}}\right) \ln \! \left(-\frac{r_{1} x_1 y_1}{u_3} \right) && \\ 
&-\text{Li}_3 \! \left(\frac{r_{1} x_{1}}{u_{3}}\right) +\text{Li}_2 \! \left(\frac{r_{1} x_{1}}{u_{3}}\right) \ln \! \left(-\frac{r_{1} x_1}{u_3} \right) - \text{Li}_3 \! \left(\frac{r_{1} y_{1}}{u_{3}}\right)+\text{Li}_2 \! \left(\frac{r_{1} y_{1}}{u_{3}}\right) \ln \! \left(-\frac{r_{1} y_{1} }{u_{3}} \right) && \\ 
& +\text{Li}_3(r_{1} x_{1}) -\text{Li}_2(r_{1} x_{1}) \ln (-r_{1} x_{1}) + \text{Li}_3(r_{1} y_{1}) - \text{Li}_2(r_{1} y_{1}) \ln (-r_{1} y_{1}) && \\ 
&-\text{Li}_3(r_{1})  +\text{Li}_2(r_{1}) \ln (-r_{1})+\frac{1}{2} \ln ^2(-r_{1}) \left[ \ln (u_3) + \ln \! \left( \frac{(1 - r_1 x_1/u_3) (1 - r_1 y_1/u_3)}{(1 - r_1 x_1)(1 - r_1 y_1) (1 - r_1 x_1 y_1/u_3)} \right) \right] && \\ 
& + \ln (-r_{1}) \left[\text{Li}_2(u_{3})  + \text{Li}_2(x_{1}) + \text{Li}_2(y_{1}) - \ln (u_{3}) \ln \! \left( \frac{(1 - r_1 x_1/u_3) (1 - r_1 y_1/u_3)}{(1 - r_1 x_1 y_1/ u_3)(1 - u_3)} \right) \right. && \\
& \quad - \ln (x_{1} y_{1}) \ln \! \left(1-\frac{r_{1} x_{1} y_{1}}{u_{3}}\right)  + \ln (x_{1}) \ln \! \left(\frac{(1- r_{1} x_{1}/u_{3}) (1 - x_1)}{1 - r_1 x_1} \right)   && \\ 
& \quad \left. + \ln (y_{1}) \ln \! \left(\frac{(1-r_{1} y_{1}/u_{3})(1 - y_1)}{1 - r_1 y_1}\right) - \frac{1}{2} \ln ^2(u_{3}) - \frac{2}{3} \pi ^2\right] \\ 
& +\frac{1}{2} \ln ^2(u_{3}) \ln \! \left(\frac{(1-r_{1} x_{1}/u_{3})(1 - r_1 y_1/u_3)}{1 - r_1 x_1 y_1/u_3} \right)   && \\ 
&   + \ln (u_{3}) \left[\ln (x_{1} y_{1}) \ln \! \left(1-\frac{r_{1} x_{1} y_{1}}{u_{3}}\right)  - \ln (x_{1}) \ln \! \left(1-\frac{r_{1} x_{1}}{u_{3}}\right) -  \ln (y_{1}) \ln \! \left(1-\frac{r_{1} y_{1}}{u_{3}}\right)\right]  && \\ 
&  - \frac{1}{2} \left(\ln^2 (x_{1} y_{1})+\pi ^2\right) \ln \! \left(1-\frac{r_{1} x_{1} y_{1}}{u_{3}}\right) + \frac{1}{2} (\ln ^2(x_{1}) + \pi^2) \ln \! \left(\frac{1 - r_1 x_1/u_3}{1-r_{1} x_{1}} \right)  && \\ 
&   + \frac{1}{2} \left (\ln ^2(y_{1}) + \pi^2 \right) \ln \! \left(\frac{1 - r_1 y_1/u_3}{1-r_{1} y_{1}}\right)-\frac{1}{6} \ln ^3(-r_{1}) +\frac{1}{2} \ln (1-r_{1}) \left(\ln ^2(-r_{1})+\pi ^2\right) && \\ 
& - \bigg \lbrace r_1 \longrightarrow r_2 \bigg \rbrace + \mathcal{O}(\epsilon) \; , && \\ 
\stepcounter{equation}\tag{\theequation}\label{eq:I11111ep}
\end{flalign*}
where we use the same notation as in Ref.~\cite{Czakon:2022fqi}:
\begin{equation}
\begin{gathered}
    x_{1} = \frac{s_{1i}}{s_{1i} + s_{2i}} \; , \qquad y_{1} = \frac{s_{2j}}{s_{1j} + s_{2j}} \; , \qquad u_{3} = \frac{s_{12} s_{ij}}{(s_{1i} + s_{2i}) (s_{1j} + s_{2j})} \; , \\
    r_{1,2} = -\frac{1}{2 x_{1}y_{1}} \left(1 - x_{1} - y_{1} - u_{3} \pm \sqrt{(1 - x_{1} - y_{1} - u_{3} )^2 - 4 x_{1} y_{1} u_{3}} \right).
\end{gathered}
\end{equation}

Let us, finally, return to analytic continuation in case of crossing. Due to rescaling invariance of the soft current, Eq.~\eqref{eq:rescaling}, the master integrals only depend non-rationally on ratios $v_1 \cdot p_{i,j}/v_2 \cdot p_{i,j}$, where $v_{1,2}$ are linear combinations of $q_1$, $q_2$, $p_i$ and $p_j$. Thus, crossing of either $p_i$ or $p_j$ alone, with both $v_{1,2}$ outgoing, cannot generate an imaginary part. On the other hand, crossing both $p_i$ and $p_j$ can generate an imaginary part due to the ratio $(v_1 \cdot p_i)(v_2 \cdot p_j)/(v_1 \cdot v_2) (p_i \cdot p_j)$, where $v_{1,2}$ are now linear combinations of $q_{1,2}$. Indeed, this ratio gains a factor of $e^{2\pi i}$ after crossing. Consequently, affected are only the master integrals $I^{(d)}_{10011}$, $I^{(d)}_{11011}$, $I^{(d)}_{10111}$, $I^{(d)}_{01111}$ and $I^{(d+2)}_{11111}$. To be specific, upon expansion in $\epsilon$, $I_{10011}^{(d)}$, $I_{11011}^{(d)}$ and $I_{10111}^{(d)}$ depend on the logarithm:
\begin{equation}
\ln \! \left( \frac{(s_{1i} + i0^+)(s_{2j} + i0^+)}{ (s_{12} + i0^+)(s_{ij} + i0^+)} \right) = \begin{cases}
\ln \! \left(\frac{s_{1i} s_{2j}}{s_{12} s_{ij}} \right) + 2 \pi i \; , \qquad &\text{for incoming Wilson lines,} \\ 
 \ln \! \left(\frac{s_{1i} s_{2j}}{s_{12} s_{ij}} \right) & \text{otherwise.}
\end{cases}
\end{equation}
$I_{01111}^{(d)}$, on the other hand, depends on the logarithm:
\begin{equation}
\ln \! \left( \frac{(s_{1i} + s_{2i} + i0^+)(s_{1j} + s_{2j} + i0^+)}{ (s_{12} + i0^+)(s_{ij} + i0^+)} \right) = \begin{cases}
\ln \! \left(\frac{(s_{1i} + s_{2i})(s_{1j} + s_{2j})}{s_{12} s_{ij}} \right) + 2 \pi i \; , \  &\text{for incoming Wilson lines,} \\ 
 \ln \! \left(\frac{(s_{1i} + s_{2i})(s_{1j} + s_{2j})}{s_{12} s_{ij}} \right) & \text{otherwise,}
\end{cases}
\label{eq:-lnZ}
\end{equation}
and further logarithms and polylogarithms resulting from the expansion of the hypergeometric function ${}_2F_1$. The analytic continuation of the latter is obtained as follows:
\begin{equation}
\begin{split}
&\text{disc} \left[ \ln (x) \right] = -2 \pi i \; , \\ 
&\text{disc} \left[ \text{Li}_2 (1 - x) \right] = 2 \pi i \ln (1 - x) \; , \\ 
& \text{disc} \left[ \text{Li}_3 (1 - x) \right] = \pi i \ln^2(1 - x) \; , \\ 
& \qquad \vdots
\end{split}
\end{equation}
where:
\begin{equation}
    x = \frac{s_{12} s_{ij}}{(s_{1i} + s_{2i})(s_{1j} + s_{2j})} \; .
\end{equation}
As far as $I_{11111}^{(d+2)}$ at $\order{\epsilon^0}$ is concerned, crossing generates an imaginary part due to the presence of $\ln(u_3)$ given by the negative of Eq.~\eqref{eq:-lnZ}. Specifically, the replacements for Eq.~\eqref{eq:I11111ep} read:
\begin{equation}
\begin{gathered}
    \ln(u_3) \; \longrightarrow \; \ln (u_3) - 2 \pi i \; , \qquad \ln \! \left( - \frac{r_{1,2} x_1 y_1}{u_3} \right) \; \longrightarrow \; \ln \! \left( - \frac{r_{1,2} x_1 y_1}{u_3} \right) + 2 \pi i \; , \\ 
    \ln \! \left(- \frac{r_{1,2} x_1}{u_3} \right) \; \longrightarrow \; \ln \! \left( -\frac{r_{1,2} x_1}{u_3} \right) + 2 \pi i  \; , \qquad \ln \! \left( - \frac{r_{1,2} y_1}{u_3} \right) \; \longrightarrow \; \ln \! \left( - \frac{r_{1,2} y_1}{u_3} \right) + 2 \pi i \; .
\end{gathered}
\end{equation}

For the currents in Eqs.~\eqref{eq:gluonJpp}, \eqref{eq:gluonJpm} and \eqref{eq:quarkJpm} the above replacements translate to:
\begin{equation}
\begin{split}
&\ln \! \left( \frac{s_{12} s_{ij}}{s_{1i} s_{2j}} \right) \; \longrightarrow \; \ln \! \left( \frac{s_{12} s_{ij}}{s_{1i} s_{2j}} \right) - 2 \pi i \; , \\ 
&\ln \! \left( \frac{s_{12} s_{ij}}{(s_{1i} + s_{2i})(s_{1j} + s_{2j})} \right) \; \longrightarrow \; \ln \! \left( \frac{s_{12} s_{ij}}{(s_{1i} + s_{2i}) (s_{1j} + s_{2j})} \right) - 2 \pi i \; , \\ 
& \ln \! \left(\frac{s_{12} s_{ij}}{s_{1i} (s_{1j} + s_{2j})} \right) \; \longrightarrow \; \ln \! \left(\frac{s_{12} s_{ij}}{s_{1i} (s_{1j} + s_{2j})} \right) - 2 \pi i \; , \\ 
& \ln \! \left( \frac{s_{12} s_{ij}}{s_{2j} (s_{1i} + s_{2i})} \right) \; \longrightarrow \; \ln \! \left( \frac{s_{12} s_{ij}}{s_{2j} (s_{1i} + s_{2i})} \right) - 2 \pi i \; , \\
&\text{Li}_2 \! \left(1 - \frac{s_{12} s_{ij}}{(s_{1i} + s_{2i})(s_{1j} + s_{2j})} \right) \; \longrightarrow \; \text{Li}_2 \! \left(1 - \frac{s_{12} s_{ij}}{(s_{1i} + s_{2i})(s_{1j} + s_{2j})} \right) \\ 
& \hspace{7cm} + 2 \pi i \ln \!  \left(1 - \frac{s_{12} s_{ij}}{(s_{1i} + s_{2i})(s_{1j} + s_{2j})} \right).
\end{split}
\end{equation}

\section{Numerical tests} \label{app:Tests}

To verify the obtained double-soft current, we performed a numerical test, in which we compared the finite part of the soft approximation with exact amplitudes of specific processes. The exact one-loop amplitudes were computed using the software package \textsc{Recola} \cite{Actis:2016mpe, Denner:2017wsf}. The software was linked to the integral library \textsc{Otter} which uses the \textsc{OpenLoops} algorithm \cite{Buccioni:2019sur}, which enabled us to use \textsc{Recola} at quadruple precision. We generated phase-space points of the Born configuration, i.e.\@ the configuration without the soft partons, using the \textsc{RAMBO} algorithm \cite{Kleiss:1985gy} which is already implemented in \textsc{Recola}. Afterwards, we applied the algorithm described in Section 2.1.2 of Ref.~\cite{Czakon:2019tmo} to generate a new phase-space point from the Born configuration. We applied the latter algorithm successively to get two soft partons with nearly the same energy. 

\begin{figure}
\begin{tabular}{cc}
    \includegraphics[scale=1]{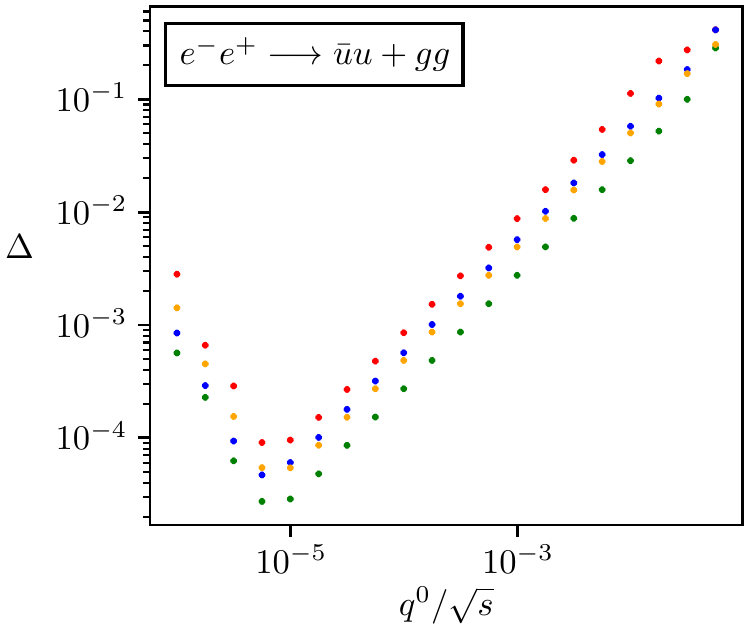}&
    \includegraphics[scale=1]{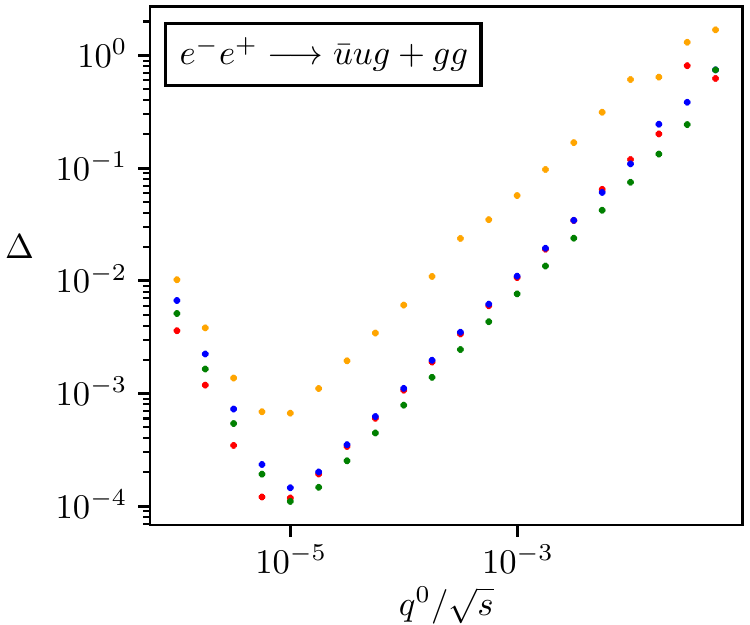}\\
    \includegraphics[scale=1]{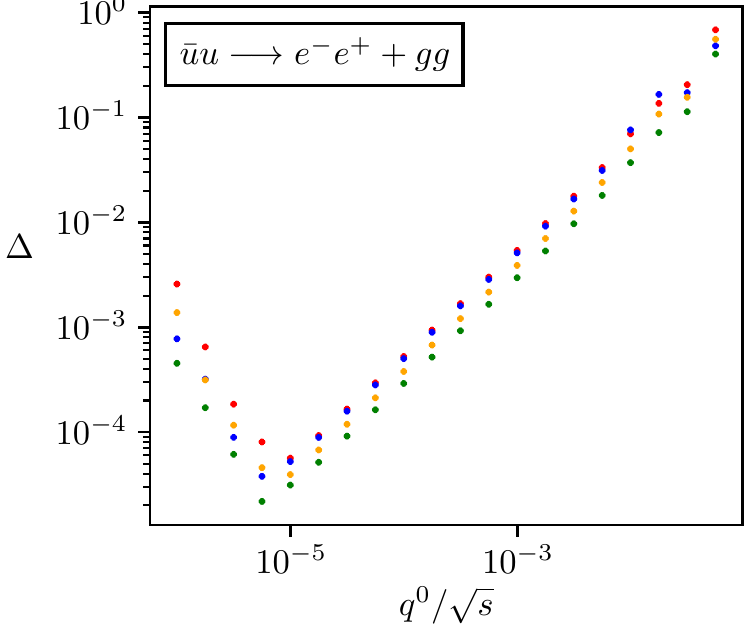}&
    \includegraphics[scale=1]{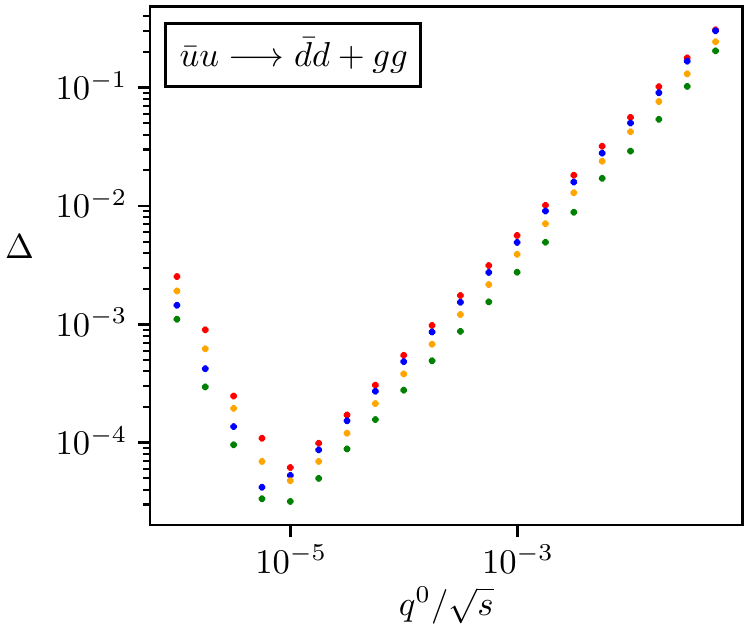}\\
    \includegraphics[scale=1]{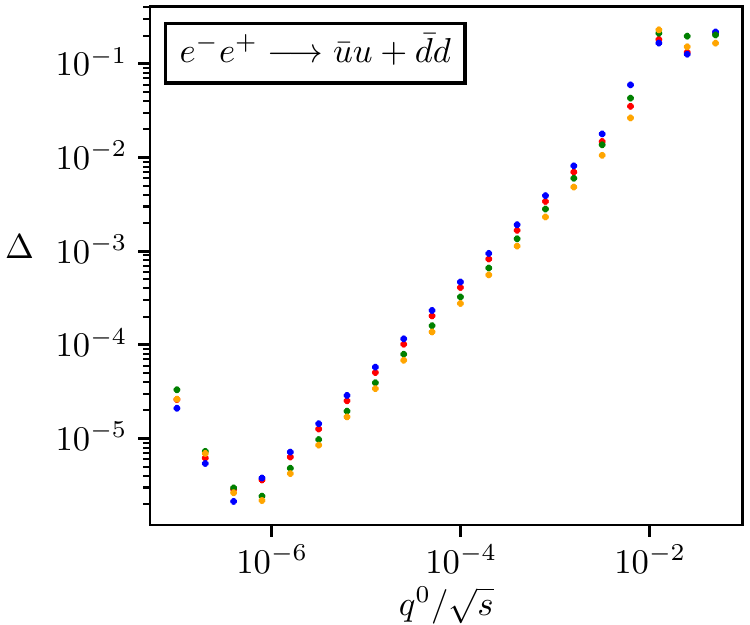}&
    \includegraphics[scale=1]{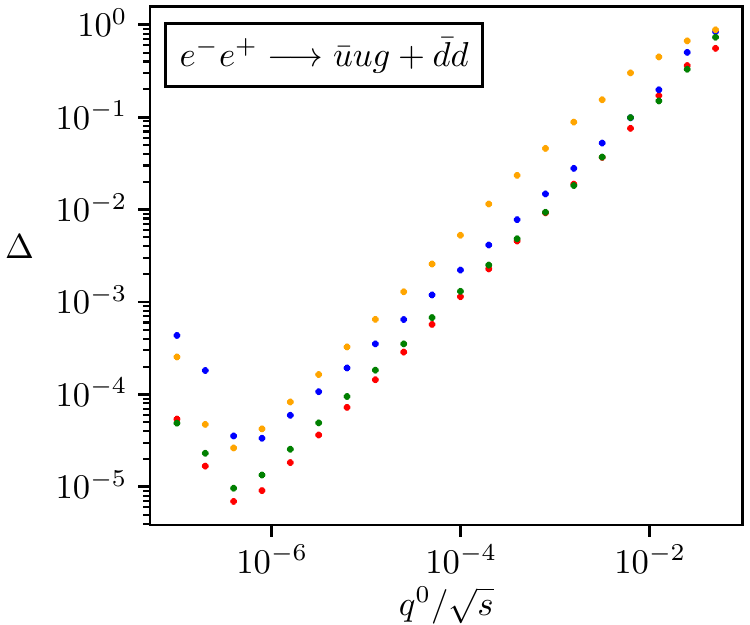}\\
\end{tabular}
\caption{Average of the relative error $\Delta$ of the one-loop double-soft approximation against the energy $q^0$ of the soft partons. The energy is normalised with the center-of-mass energy $\sqrt{s}$ of the process. The different colours correspond to different Born phase-space points.}
\label{fig:numerical-test}
\end{figure}

Fig.\@ \ref{fig:numerical-test} shows the average of the relative error for different values of $q_1^0 \approx q_2^0 \equiv q^0$ of the eikonal approximation:
\begin{equation}
\Delta = \frac{1}{N} \sum_{\substack{\text{leading power} \\ \text{colour, spin}}} \bigg \vert \frac{\bra{a_1 \lambda_1, \ldots , a_{n + 2} \lambda_{n + 2}} \ket{ M_{n + 2}} - \bra{a_1 \lambda_1, \ldots , a_n \lambda_n } \ket{J^{a_{n+1} a_{n+2}} (q_1, q_2) | M_n} }{\bra{a_1 \lambda_1, \ldots , a_{n + 2} \lambda_{n + 2}} \ket{  M_{n + 2}}} \bigg \vert \; ,
\end{equation}
where only the finite remainder of the one-loop contribution in the numerator and denominator is kept. The summation is performed over all spin and colour configurations with soft singularities. Configurations which result in non-singular amplitudes are excluded since the soft approximation does not apply to them. The different data sets were computed using different Born phase-space points.

Fig.\@ \ref{fig:numerical-test} clearly shows the expected $\Delta\ \propto \ q^0$
dependence of a leading-power approximation up until about $q^0/\sqrt{s} = 10^{-5}$ for gluons and $q^0/\sqrt{s} = 10^{-6}$ for quarks, where the numerical precision of the integral computation in \textsc{Recola/Otter} is no longer sufficient.

\section{\texorpdfstring{Erratum for Ref.~\cite{Zhu:2020ftr}}{Erratum for Ref.~[1]}} \label{app:Erratum}

\begin{itemize}
    \item The Ward identities in Eq.\@ (2.10) should include a minus sign in front of the single-soft current (compare to our Eq.~\eqref{eq:2gluonJ0Ward}).
    \item The result for the integral $I_{11011}$ in Eq.\@ (4.8) should be divided by 4.
    \item The expression for the integral $I_{11111}$ in Eq.\@ (4.9) should contain the second power of the logarithm: 
    \begin{equation}
        \ln \! \left( \frac{s_{14} s_{23}}{(s_{12} + s_{13}) s_{34}} \right) \; ,
    \end{equation}
    at $\order{\epsilon^0}$, instead of the first power. This is obvious since the result is of uniform transcendentality.
    \item Terms proportional to the colour structure $ i f_{b a_2 a_3} \mathbf{T}_i^b$ in Eq.~(3.6) differ by a minus sign from our own results \eqref{eq:gluonJppHat} (see discussion below Eq.~\eqref{eq:gluonJpm}). 
    \item Results for soft quark-anti-quark emission differ in sign from ours and those of Ref.~\cite{Catani:2021kcy}, as the author of Ref.~\cite{Zhu:2020ftr} acknowledges below Eq.~(3.22). However, unlike in the single-soft case, such a sign difference cannot be explained by a different convention for the sign of the strong coupling constant, contrary to what is claimed in Ref.~\cite{Zhu:2020ftr}.
\end{itemize} 

\section{Description of ancillary files} \label{app:Files}

The following files, besides the last, are provided in \textsc{Mathematica} format:
\begin{itemize}
    \item \texttt{J\_gg.m} -- Double soft-gluon emission soft current with exact $\epsilon$-dependence in terms of master integrals, see Section~\ref{sec:ExactResults}.
    \item \texttt{J\_gg\_ep0.m} -- Double soft-gluon emission soft current accurate at $\mathcal{O}(\epsilon^0)$. The expansion is missing the factor $r_\Gamma \left(\frac{-s_{12} - i 0^+}{\mu^2} \right)^{-\epsilon}$, see Section~\ref{sec:ExpandedResults}.
    \item \texttt{J\_qq.m}, \texttt{J\_qq\_ep0.m} -- Analogous files to the two previous ones for soft quark-anti-quark pair emission, see Section~\ref{sec:SoftQuarks}.
    \item \texttt{integrals.m} -- Exact results from Appendix~\ref{app:Integrals} for all integrals except $I_{11111}^{(d+2)}$ as a replacement list. The latter list can be used together with \texttt{J\_gg.m} to reproduce \texttt{J\_gg\_ep0.m}, and analogously for \texttt{J\_qq.m} and \texttt{J\_qq\_ep0.m}.
    \item \texttt{I11111.m} -- Series expansion of $I_{11111}^{(d+2)}$ in $\epsilon=(4-d)/2$ up to $\mathcal{O}(\epsilon^1)$ reproduced from Ref.~\cite{Czakon:2022fqi}. This result can be used to obtain soft currents accurate at $\mathcal{O}(\epsilon^2)$.
    \item \texttt{README} -- Explanation of files and used notation.
\end{itemize}

\newpage

\bibliographystyle{JHEP}
\bibliography{main} 

\end{document}